\documentclass[aps,twocolumn,pre]{revtex4-1}
\usepackage{graphicx,amsmath, amsfonts,bm, color}

\usepackage{xcolor}
\usepackage[hypertexnames=false]{hyperref}
\hypersetup{
	colorlinks,
	linkcolor={blue!50!black},
	citecolor={blue!50!black},
	urlcolor={blue!80!black}
}

\usepackage{bm}
\usepackage{braket} 

\usepackage{amsfonts,amssymb}
\renewcommand{\L}{\mathfrak{L}}
\newcommand{\D}{\mathfrak{D}}
\newcommand{\1}{^{(1)}}
\newcommand{\2}{^{(2)}}
\newcommand{\pa}{\partial}
\newcommand{\DT}{^{\mbox{\tiny DT}}}
\newcommand{\SFT}{^{\mbox{\tiny CT}}}
\newcommand{\uc}{_{\mbox{\scriptsize uc}}}
\newcommand{\erf}{\operatorname{erf}}
\newcommand{\eff}{_{\mbox{\scriptsize eff}}}
\newcommand{\sint}{_{\mbox{\scriptsize int}}}
\newcommand{\pd}[2]{\frac{\partial #1}{\partial #2}}
\newcommand{\pdd}[2]{\frac{\partial^2 #1}{\partial #2^2}}

\begin{document}
\title{Gaussian fluctuations of spatially inhomogeneous polymers}

\author{Yohai Bar-Sinai}
\thanks{Current address: School of Engineering and Applied Sciences, Harvard University, 29 Oxford St., Cambridge, MA 02138, USA.}
\email[E-mail: ]{ybarsinai@gmail.com}

\author{Eran Bouchbinder}
\email[E-mail: ]{bouchbinder@weizmann.ac.il}
\affiliation{Chemical Physics Department, Weizmann Institute of Science, Rehovot 7610001, Israel.}

\begin{abstract}
Inhomogeneous polymers, such as partially ligand-bound DNA or partially cofilin-bound actin filaments, play an important role in various cellular processes, both in nature and in biotechnological/therapeutic applications. At finite temperatures, inhomogeneous polymers exhibit non-trivial thermal fluctuations. In a broader context, these are relatively simple examples for fluctuations in spatially inhomogeneous systems, which are less understood compared to their homogeneous counterparts.
Here we develop a statistical theory of torsional, extensional and bending Gaussian fluctuations of inhomogeneous polymers (chains), where the inhomogeneity takes the form of an inclusion of variable size and mechanical properties, using both continuum and discrete approaches. First, we analytically calculate the complete eigenvalues and eigenmodes spectrum of the inhomogeneous polymer within a continuum field theory. In particular, we show that the wavenumber inside and outside of the inclusion is nearly linear in the eigenvalue index, with a nontrivial coefficient. Second, we solve the corresponding discrete problem, and highlight fundamental differences between the continuum and discrete spectra of eigenvalues/eigenmodes. In particular, we demonstrate that above a certain wavenumber the discrete spectrum changes qualitatively and discrete evanescent eigenmodes, that do not have continuum counterparts, emerge. The statistical thermodynamic implications of these differences are then explored by calculating fluctuation-induced forces associated with free-energy variations with either the properties of the inclusion (e.g.~inhomogeneity formed by adsorbing molecules) or with an external geometric constraint. The former, which is the fluctuation-induced contribution to the adsorbing molecules binding force, is shown to be affected by short wavelengths and thus cannot be calculated using the continuum approach. The latter, on the other hand, is shown to be dominated by long wavelength shape fluctuations and hence is properly described by the continuum theory.
\end{abstract}

\maketitle

\section{Introduction}
\label{sec:Intro}

Spatially inhomogeneous systems are ubiquitous in the natural and manmade world around us, giving rise to intriguing physical behaviors as compared to their homogeneous counterparts. For example, glassy systems --- which feature inhomogeneity/disorder on small lengthscales --- still pose great challenges in condensed-matter and statistical physics~\cite{Ediger2000, Cavagna2009, Berthier2011}. Low-dimensional systems, such as rods, beams and polymers, also feature interesting behaviors in the presence of spatial inhomogeneities in their properties. When these systems are excited externally, either by mechanical perturbations or by coupling to a heat bath, they exhibit non-trivial responses and fluctuations associated with the spatial inhomogeneity. Thermal and entropic effects are known to play a major role in a broad range of soft matter and biophysics problems where polymers and biopolymers are considered~\cite{deGennes1979, Safran1994}. Therefore, it is important to understand the effect of spatial inhomogeneity on the fluctuations of polymers~\cite{Helfand1975, Popov2007, Su2010}.

To address this problem we study in this paper the mechanics and statistical thermodynamics of spatially inhomogeneous one-dimensional polymers where the inhomogeneity takes the form of an inclusion of finite length which is mechanically softer than the rest of the polymer. The polymer is assumed to be submerged in a solvent of a fixed temperature such that it undergoes overdamped equilibrium thermal fluctuations under certain constraints. There are many physical systems that might give rise to such a situation. For example, actin filaments in cells are known to significantly soften in regions where cofilin molecules bind to them~\cite{Prochniewicz2005, McCullough2008, Yogurtcu2012, Galkin2011}, so partially cofilin-decorated actin filaments are spatially inhomogeneous. Other natural and man-made systems can exhibit similar spatial inhomogeneity~\cite{Helfand1975, Panyukov1996, Bensimon1998, Lipfert2010, Kostjukov2012}.

Our discussion, while motivated by these realistic and important examples, remains rather general and independent of the particular details of the underlying physical system. This is achieved by considering general Hamiltonians in the small gradient approximation, i.e. generic quadratic Hamiltonians (energy functionals). In the context of torsional fluctuations of the polymer, the energy functional is quadratic in the gradient of the twist angle, while in the context of extensional fluctuations of the polymer the energy functional is quadratic in the gradient of the longitudinal displacement along the polymer. The resulting energy functional is the same in these two cases.

In the context of bending fluctuations of the polymer, the energy functional is quadratic in the the local curvature, which in itself is a second derivative of the out-of-plane deflection of the polymer in the small gradient approximation. This is nothing but the classical one-dimensional Helfrich Hamiltonian in the absence of surface tension~\cite{Helfrich1973, Safran1994}. As such, from a more theoretical perspective, we consider the classical example of massless\footnote{In the jargon of statistical field theory, ``massless'' refers to the lack of a term proportional to $w^2$, which in our case would correspond to an external potential. Here we also work in the overdamped limit, which in this context might be termed ``intertia-less''.} quadratic field theory in one spatial dimension with position-dependent properties, applicable to a broad range of other physical systems~\cite{KardarFields}.

Gaussian fluctuations of such one-dimensional fields, i.e. when the quadratic approximation is adopted, are oftentimes addressed in the framework of statistical field theory. In this framework a continuum approach is invoked and macroscopic variables of interest are assumed to vary slowly in space. One of our goals here is to understand to what extent the problem can be described by the continuum approach and when does it break down. To that aim, we solve the problem using both a continuum field theory and its discrete counterpart.

We highlight the fundamental differences between the continuum and discrete spectra of eigenvalues and eigenmodes, and explore the implications of these differences in relation to two physically realistic fluctuation-induced forces. The first one is a fluctuation-induced force associated with free-energy variations with respect to the properties of the inclusion (e.g.~formed by adsorbing molecules, in which case it is the fluctuation-induced contribution to the adsorbing molecules binding force), while the second is a fluctuation-induced force associated with free-energy variations with respect to an external geometric constraint (e.g.~a confining wall). We show that while the continuum theory is valid in the latter case, it breaks down in the former.

\section{Mathematical formulation}
\label{sec:math}

We consider a spatially inhomogeneous one-dimensional polymer of length $L$, consisting of $N$ monomeric units, submerged in a solvent of temperature $T$. $x\in[0,L]$ is the coordinate along the polymer. The inhomogeneous polymer is treated at the continuum level as a one-dimensional beam/rod characterized by position-dependent mechanical properties along its axis $x$. The polymer's length is assumed to be comparable to its persistence length with respect to torsional, extensional and bending fluctuations, which implies that the polymer is semi-flexible and hence is fully characterized by its elastic energies.

For concreteness, we consider a polymer composed of $3$ locally homogeneous regions with sharp interfaces between them. In other words, we consider an inclusion inside a polymer such that the space-dependent elastic modulus reads
\begin{equation}
 \kappa(x)=\begin{cases}
            \kappa_s & x_1<x<x_2\\
            \kappa_h & x<x_1\ \mbox{ or }\ x>x_2
           \end{cases}\ .
\end{equation}
That is, $\kappa$ equals $\kappa_s$ inside the inclusion and $\kappa_h$ otherwise. The subscripts $h,s$ denote ``hard'' and ``soft'', respectively, so that $\kappa_h\!>\!\kappa_s$. Here, $\kappa$ refers generically to either of the torsional, extensional or bending moduli.

We consider the small gradient approximation in which the torsional, extensional and bending are described by quadratic energy functionals. At the continuum level, this leads to Gaussian fluctuations that are controlled by either of the two quadratic energy functionals
\begin{equation}
\begin{aligned}
 U\1(x,t)&=\frac{1}{2}\int _0 ^L  \kappa(x)\left(\pd{w(x,t)}{x}\right)^2\ dx \ , \\
 U\2(x,t)&=\frac{1}{2}\int _0 ^L  \kappa(x)\left(\pdd{w(x,t)}{x}\right)^2\ dx \ ,
\end{aligned}
 \label{eq:uc}
\end{equation}
where $w(x,t)$ is a fluctuating field and $\kappa(x)$ is its related modulus. Torsional and extensional fluctuations are described by the former, while bending fluctuations are described by the latter. In torsional dynamics $w(x)$ represents the twist angle, measured relative to an a priori given equilibrium twist angle profile $\theta_0(x)$~\cite{McCullough2011}. In extensional dynamics, $w(x)$ measures the longitudinal displacement along the axis of the polymer. In bending dynamics, $w(x)$ measures the normal deviation of the polymer from a straight line, i.e.~the out-of-plane deflection (we assume that the polymer does not feature any intrinsic curvature). This is nothing but the classical one-dimensional Helfrich Hamiltonian in the absence of surface tension~\cite{Helfrich1973, Safran1994}.

The very same problem can be formulated at the discrete level, making reference to monomeric degrees of freedom and lengthscales. In particular, the discrete analogs of Eqs.~\eqref{eq:uc} take the form
\begin{equation}
\begin{aligned}
U\1&=\frac{1}{2}\sum_{i} \kappa_i\left(\frac{w_i-w_{i-1}}{a}\right)^2a\ , \\
U\2&=\frac{1}{2}\sum_{i} \kappa_i\left(\frac{w_{i+1}+w_{i-1}-2w_i}{a^2}\right)^2a\ ,
\end{aligned}
 \label{eq:ud}
\end{equation}
where $a\!\equiv\!L/N$ is a monomeric lengthscale, $\kappa_i$ and $w_i$ are the discrete version of $\kappa(x)$ and $w(x)$, respectively, and $i$ is the monomer index. 

Equations~\eqref{eq:uc}-\eqref{eq:ud} are representative of a wide class of physical systems whose energy functionals, in the quadratic approximation, can be written as
\begin{equation}
\begin{aligned}
  U&=\frac{1}{2}\braket{w(x)|\L|w(x)}=\frac{1}{2L}\int_{0}^{L}\!\! w(x)\,\L\{w(x)\}\,dx\ ,\\
  U&= \frac{1}{2}\, \braket{\bm w|\bm H|\bm w}= \frac{1}{2}\sum_{i,j} w_i H_{ij} w_j\ ,
\end{aligned}
 \label{eq:u_general}
\end{equation}
where $\bm H$ is a real symmetric positive definite matrix and $\L$ is a self-adjoint real differential operator~\cite{Arfken}. Equations~\eqref{eq:uc}-\eqref{eq:ud} are recovered from Eqs.~\eqref{eq:u_general} with the proper identification of the dynamical operator. For the discrete theory, one clearly has $\bm H\!=\!\bm{\nabla\nabla} U$. For the continuum theory, one finds that the dynamical operators related to $U\1$ and $U\2$ are respectively
\begin{equation}
\begin{aligned}
 \L\1\{w\}& =-L\pd{}{x}\left(\kappa(x)\pd{w}{x}\right)  & \mbox{for } U\1 \ ,\\
 \L\2\{w\}&=-L\pdd{}{x}\left(\kappa(x)\pdd{w}{x}\right) & \mbox{for } U\2 \ .
\end{aligned}
\end{equation}

We adopt here the convention that the eigenvalues of $\L$ or $\bm H$ are of energy dimensions, and accordingly the variables $w_i$ and $w(x)$ are dimensionless. In addition, it would be useful to introduce the dimensionless parameters $\phi$ and $\Delta$,
\begin{align}
\label{eq:phi_delta}
    &\phi\equiv\frac{x_2-x_1}{L}\ ,\\
    &\Delta\!\equiv\!\left(\frac{\kappa_h}{\kappa_s}\right)^{1/2} \quad\mbox{for}\quad U\1 ,\quad
    \Delta\!\equiv\!\left(\frac{\kappa_h}{\kappa_s}\right)^{1/4} \quad\mbox{for} \quad U\2\ ,\nonumber
\end{align}
which are measures of the inclusion size and contrast, respectively. Note that $0\!\le\!\phi\!\le\!1$ and $\Delta\!\ge\!1$. Finally, in order to completely define the problem one needs to specify also the external boundary conditions at $x\!=\!0,L$. Here we take the polymer to be fixed (pinned) at $x\!=\!0$ and free at $x\!=\!L$. Mathematically, this means
\begin{equation}
\begin{aligned}
 w(0)&\!=\!w'(L)\!=\!0 								&\mbox{for } U\1,\\
 w(0)&\!=\!w'(0)\!=\!w''(L)\!=\!w'''(L)\!=\!0 \quad&\mbox{for }U\2,
\end{aligned}
\end{equation}
where a prime denotes partial differentiation with respect to $x$. For the discrete formulation, this amounts to setting $w_i\!=\!0$ for $i\!<\!1$ and $\kappa_i\!=\!0$ for $i\!>\!N$. Choosing different boundary conditions does not qualitatively change the results presented below.

\section{Eigenmode analysis: Continuum theory}
\label{sec:CT}

Gaussian fluctuations are fully determined by the eigenvalues of the relevant dynamical operator. Consequently, an essential step in the statistical thermodynamic calculations to follow is finding the eigenvalues and the corresponding eigenmodes of $\L$ or $\bm H$. This will be the subject of this section and the next one. In this section we calculate the eigenmodes within the continuum theory, and show that the wavenumbers have a constant density. In Sec.~\ref{sec:DT} the corresponding discrete problem is solved and the differences between the results are discussed.

\subsection{General form of the eigenmodes}
\label{sec:continuum}

The calculation of the eigenmodes is very much in the spirit of standard wave theory analysis of reflection and refraction from a sharp material boundary, or of the quantum mechanical treatment of transmission over a potential barrier step. The eigenmodes $w_q$ are functions that satisfy the continuum eigenvalue equation --- the Sturm-Liouville problem---,
\begin{equation}
 \L w_q(x)=\lambda_q w_q(x)\ ,
 \label{eq:eigen}
\end{equation}
where $\lambda_q$ is the eigenvalue associated with $w_q$. Solving Eq.~\eqref{eq:eigen} is in general a non-trivial task. However, since $\kappa(x)$ is locally constant for $x\!\ne\!x_1,x_2$, treating the soft and hard polymeric segments separately significantly simplifies the mathematical structure. That is, in each segment $\kappa$ is space-independent, such that except at the discontinuity points, Eq.~\eqref{eq:eigen} reads

\begin{align}
 \label{eq:spatial1}
\kappa(x)\, w_q''(x)&=\lambda_q\, w_q(x)   \ , &\mbox{for} \quad\L\1\ ,\\
 \label{eq:spatial2}
\kappa(x)\, w_q''''(x)&=\lambda_q\, w_q(x) \ , &\mbox{for} \quad\L\2\ .
\end{align}

It is thus natural to write the solution separately for the different segments. For each segment, we write $w_q(x)$ as a superposition of the independent solutions of Eqs.~\eqref{eq:spatial1}-\eqref{eq:spatial2}. For $\L\1$, the solution of Eq.~\eqref{eq:spatial1} reads
\newcommand{\tq}{\tilde{q}}
\begin{equation}
\label{eq:eigenmodes}
 w_q(x)\!=\!\begin{cases}
      A_1 \cos(q x) + A_2 \sin(q x) & 0_{\hphantom 1}<x<x_1\\
      A_3 \cos(\tq x) + A_4 \sin(\tq x) & x_1<x<x_2\\
      A_5 \cos(q x) + A_6 \sin(q x) & x_2<x<L
     \end{cases}\ ,
\end{equation}
where the $A_i$ are yet undetermined real amplitudes. We also impose the supplementary condition
\begin{align}
 q^2 \kappa_h=\tq^2 \kappa_s \qquad  \mbox{or equivalently} \qquad \tilde q=\Delta\,q\ ,
\label{eq:dispersion_1}
\end{align}
which ensures that Eq.~\eqref{eq:spatial1} is satisfied with the same eigenvalue $\lambda_q\!=\!L\,\kappa_h \,q^2\!=\!L\,\kappa_s \,\tq^2$ at all points in space. A mode with negative $q$ can be obtained by rearranging the coefficients $\{A_i\}$ in the corresponding mode with a positive $q$, so we only consider modes with $q\!>\!0$.
Similarly, for $\L\2$ we write the solution $w_q(x)$ of Eq.~\eqref{eq:spatial2} as a combination of $\cos(q x)$, $\sin(q x)$, $\cosh(q x)$ and $\sinh(q x)$, with the supplementary condition
\begin{equation}
 q^4 \kappa_h=\tq^4 \kappa_s \qquad  \mbox{or equivalently} \qquad \tilde q=\Delta\,q \ .
\label{eq:dispersion_2}
\end{equation}
Note that $\Delta$ is defined differently for the two operators, cf.~Eq.~\eqref{eq:phi_delta}.

\begin{figure*}
 \centering
 \includegraphics[width=\textwidth]{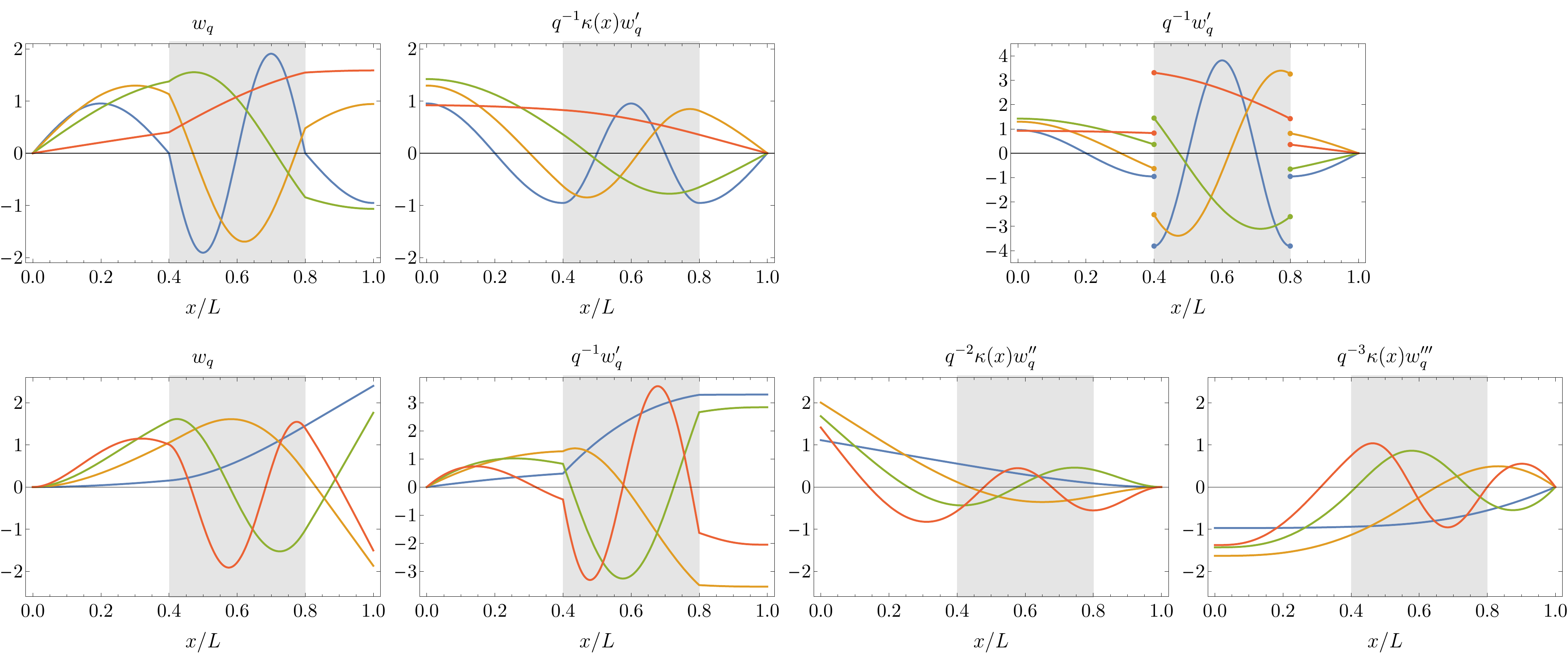}
 \caption{Lowest 4 modes of $\L\1$ (top row) and $\L\2$ (bottom row) and their derivatives. The leftmost panels show the modes and successive panels show successive derivatives. In cases where the derivatives are discontinuous we plot the derivatives multiplied by the discontinuous $\kappa(x)$, which results in continuous functions, cf.~Eqs.~\eqref{eq:jumps_1}-\eqref{eq:jumps_2}. As an example, the rightmost panel in the top row shows $w'$ itself, explicitly demonstrating the discontinuity. The parameters used here and in what follows are $x_1\!=\!0.4L$, $x_2\!=\!0.8L$ and $\Delta\!=\!1.5$. The shaded area shows the region in space where $\kappa(x)\!=\!\kappa_s$ and it is readily seen that in this region the wavelength of the modes is shorter.}
 \label{fig:modes}
\end{figure*}

\subsection{Internal boundary conditions}

A crucial step in calculating the structure of the eigenmodes is specifying the internal boundary conditions (BC) at the discontinuity points $x\!=\!x_1,x_2$. These, together with the external boundary conditions at $x\!=\!0,L$, determine the amplitudes $\{A_i\}$. It is important to stress that the external boundary conditions completely and uniquely specify the Sturm-Liouville problem. However, since we treat the problem separately for the different segments, we also need to specify the \textit{internal} BC at the mechanical discontinuity points. That is, the internal BC are a result of our choice to divide the problem into 3 distinct segments. If $\kappa$ were to change over a finite length-scale, then this division would not have been necessary (nor possible) and no internal BC would have been needed. Such a calculation is carried out in the supplementary material~\cite{SM}, though in this case it cannot be carried out analytically.

The form of the internal BC can be obtained either by taking the limit of an infinitely small variation length of $\kappa$, or equivalently, in the following manner. The spatiotemporal dynamics of the system are governed by the equation
\begin{equation}
 \mathcal{T}\{w(x,t)\}=\L\{w(x,t)\}\ ,
 \label{eq:t_operator}  
\end{equation}
where $\mathcal{T}$ is a differential operator acting on the time coordinate. $\mathcal{T}\{w(x,t)\}$ is proportional to $\partial_{tt}w(x,t)$ in inertial systems, to $\partial_t w(x,t)$ in highly overdamped systems and might have a more complicated structure in other cases. Since the particular form of $\mathcal{T}$ is irrelevant to this discussion, we do not specify it here. We integrate Eq.~\eqref{eq:t_operator} over a region of size $\delta$ around the a discontinuity point, say $x_1$. That is, we consider the region  $-\delta\!<\!x-x_1\!<\!\delta$ and take the limit $\delta\to0$. Using the fact that for $x\!\ne\! x_1$ $\kappa$ is space-independent, the integration can be done explicitly. For $\L\1$ the result is
\begin{align*}
\begin{split}
 \lim_{\delta\to\,0}\int_{x_1-\delta}^{x_1+\delta}&\mathcal{T}\{w(x,t)\} dx =\\
 &-L\lim_{\delta\to\,0}\Big[\kappa_s\left.w'\right|_{x=\delta}-\kappa_h\left.w'\right|_{x=-\delta}\Big]\ .
\end{split}
\end{align*}
Irrespective of the explicit form of $\mathcal{T}$, we know that it does not produce a singularity at $x\!=\!x_1$ and thus the left-hand-side of the above equation vanishes. We therefore conclude that the function $\kappa(x)\,w'\!(x)$ is continuous across the interface. Repeating this procedure again shows that $w(x)$ is continuous across $x=x_1$. As before, one uses the fact that although $\kappa$ is discontinuous, it is not singular and its integral over a vanishingly small region vanishes.

To summarize, the 4 internal boundary conditions for $\L\1$ are
\begin{align}
  [\![w]\!]_{x_1}\!=[\![w]\!]_{x_2}\!=[\![\kappa w']\!]_{x_1}=\![\![\kappa w']\!]_{x_2}=0\ ,
  \label{eq:jumps_1}
\end{align}
where $[\![\cdot]\!]_{x_i}$ denotes the jump of a given quantity at $x=x_i$. In particular, as $\kappa(x)$ is discontinuous at $x_1$ and $x_2$, $w'\!(x)$ experiences a jump-discontinuity at these points. The somewhat formal derivation of the internal BC at $x_1$ and $x_2$ presented above has a clear physical meaning that could have been invoked a priori; at any discontinuity of the linear elastic modulus $\kappa$, the polymer retains its integrity, i.e. $w(x)$ is continuous, and the stress (either torsional or extensional) is continuous, i.e. $\kappa(x)\,w'\!(x)$ is continuous.

Similarly, for $\L\2$ one obtains that the internal boundary conditions at the discontinuity points are
\begin{align}
  [\![w]\!]=[\![w']\!]=[\![\kappa w'']\!]=[\![\kappa w''']\!]=0\ .
  \label{eq:jumps_2}
\end{align}
The last two conditions physically correspond to continuity of the mechanical torque and shear force in the polymer.
\begin{figure*}
 \centering
  \includegraphics[width=\textwidth]{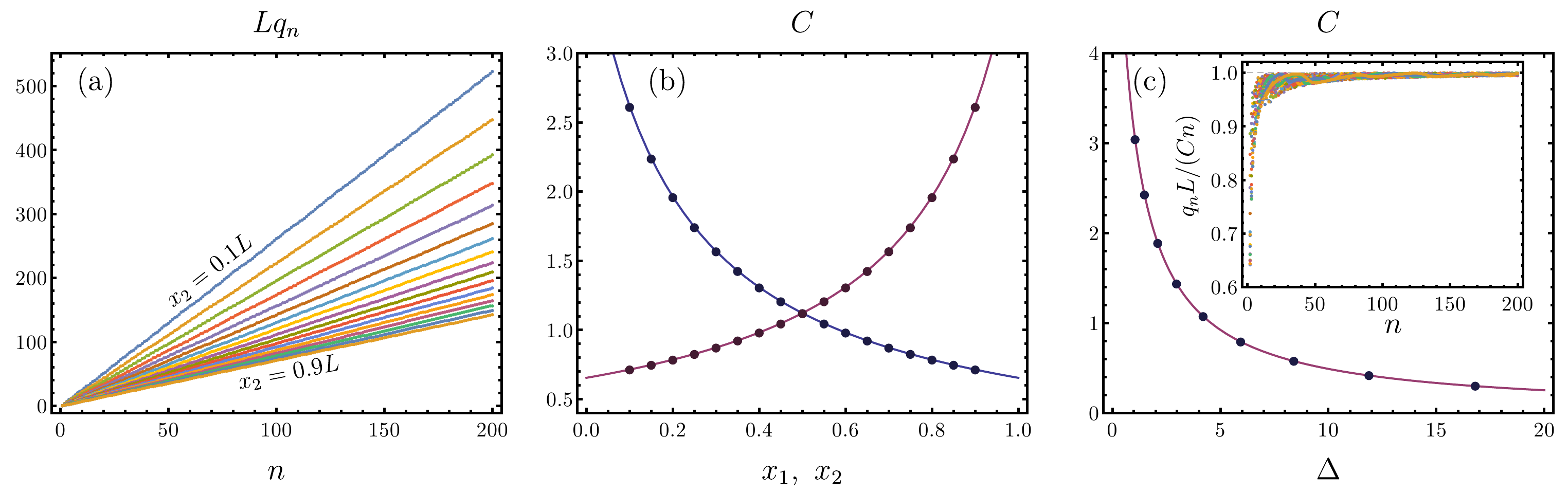}
 \caption{\textbf{(a)} The numerically found $q$'s of $\L\1$ as a function of their ordinal number for fixed $x_1\!=\!0.05L$ and $\Delta\!=\!5$. Different colors correspond to different values of $x_2$ which varies at constant steps between $0.1L$ and $0.9L$. \textbf{(b)} The blue points show the slopes of the data in panel (a) as a function of $x_2$, and the solid line is the prediction of Eq.~\eqref{eq:slopes}. The purple data are obtained with the same procedure, but when $x_2=0.95L$ is fixed and $x_1$ varies. \textbf{(c)} The same as (b), but now $\Delta$ is varied and $x_1\!=\!0.2L$, $x_2\!=\!0.8L$ are fixed. Inset: the same data and color code as in panel (a), normalized by the predicted value $Cn$. It is seen that the ratio exhibits significant deviations from unity only for the first few modes.}
 \label{fig:dispersion}
\end{figure*}

\subsection{The spectrum of permissible wavenumbers $q_n$}

The boundary conditions specified above are all linear and therefore can be summarized concisely in a matrix equation
\begin{equation}
\bm M(q; \Delta, x_1,x_2)
\vec A=0
 \ ,
\end{equation}
where $\vec A$ is the vector of amplitudes and $\bm M$ is a matrix which can be explicitly calculated. In order to satisfy the boundary conditions simultaneously one must demand $\det\bm M=0$. The resulting equation can be solved numerically to find the discrete set of permissible $q$'s. For each permissible $q$, the eigenvectors are easily found by calculating the kernel of the matrix.

For example, the equation that defines the permissible $q$'s for $\L\1$ explicitly reads
 \begin{align}
\nonumber  0=&\hphantom{+}\left(\frac{\Delta +1 }{\Delta -1}\right)^2\cos\Big[q ((x_2-x_1) (\Delta -1)+L)\Big]\\
\nonumber  &+\frac{(\Delta +1) }{(\Delta -1)}\cos\Big[q ((x_2-x_1) \Delta -(x_2+x_1)+L)\Big]\\
\nonumber  &-\frac{(\Delta +1) }{(\Delta -1)}\cos\Big[q (-(x_2-x_1) \Delta -(x_2+x_1)+L)\Big]\\
  &-\cos\Big[q ((x_1-x_2) (1+\Delta )+L)\Big]\ .
 \label{eq:dispersion}
 \end{align}
This equation can be solved numerically and the first few modes of $\L\1$ and $\L\2$ are shown in Fig.~\ref{fig:modes} and briefly discussed in its caption.

The challenge now is to estimate how the permissible $q$'s are distributed as a function of the parameters. To this end, we numerically solve Eq.~\eqref{eq:dispersion} for some range of inclusion parameters. In Fig.~\ref{fig:dispersion} we plot the numerically found wavenumbers $\{q_n\}$ as a function of their ordinal number $n$, when $x_2$ is varied while $x_1$ and $\Delta$ are fixed. It is seen that for each fixed set of parameters the spectrum is quasilinear, i.e.~that one can approximately write the $n$-th wave number as
\begin{equation}
\label{eq:DOS}
 q_n \approx \frac{C(\Delta,x_1,x_2)}{L}n\ .
\end{equation}
Simple dimensional analysis of Eq.~\eqref{eq:dispersion} shows that $C$ cannot depend on $L$ nor on $\kappa_h$ or $\kappa_s$, except through their ratio $\Delta^2$.

How can we estimate $C(\Delta,x_1,x_2)$? The defining equation, Eq.~\eqref{eq:dispersion}, is a sum of sinusoidal functions with different frequencies. One can conjecture that the highest frequency, $(\Delta -1) (x_2-x_1)+L$, is the one that controls the density of solutions. This argument predicts that the equation for $C$ should read
\begin{equation}
\label{eq:slopes}
 C\simeq\frac{\pi L}{(\Delta-1)(x_2-x_1)+L}=\frac{\pi}{\phi\Delta+1-\phi}\ .
\end{equation}
Figure~\ref{fig:dispersion} demonstrates a numerical verification of this prediction. Note that $C$ depends on $x_1$ and $x_2$ only through their normalized difference $\phi$, i.e.~that $C$ does not depend on the location of the inclusion but only on its relative size. In fact, the same relation holds also if two or more inclusions are present. In this case $C$ depends on the total fraction of the polymer which is occupied by the inclusions (not shown). For the operator $\L\2$, the analysis is similar yet more technically involved. The final result, though, is identical --- the wavenumber of the $n$-th eigenmode is quasilinear in $n$ and the proportionality factor is given by Eq.~\eqref{eq:slopes} (although the definition of $\Delta$ is different, cf.~Eq.~\eqref{eq:phi_delta}).

The constant $C$ provides a closed-form, non-perturbative approximation for the structure of the spectrum of $\L$. It can also be derived heuristically with the following reasoning. Writing the denominator of $C$ as $\phi\cdot\Delta+(1-\phi)\cdot1$, it is seen that it is a rule of mixture between $\Delta$ and 1, with relative weights of $\phi$ and $1-\phi$, respectively. In the spirit of Eqs.~\eqref{eq:eigenmodes}-\eqref{eq:dispersion_2}, an eigenmode of either $\L\1$ or $\L\2$ can be written schematically as
\begin{equation}
 w_q(x)\sim\begin{cases}e^{iqx} & \mbox{in the stiff regions}\\ e^{iq \Delta x} & \mbox{in the soft regions}\end{cases} \ .
\end{equation}
Thus, if we ``stretch'' the $x$ coordinate in the softer regions by an amount $\Delta$, the eigenmode will have the same wavenumber in both regions. That is, if we define a new variable $\tilde x$ by the differential $d\tilde x\!\equiv\!\sqrt{\kappa_h/\kappa(x)}\,dx$ for $\L\1$ and $d\tilde x\!\equiv\!\sqrt[4]{\kappa_h/\kappa(x)}\,dx$ for $\L\2$,
then $w(\tilde x)$ has the same wavenumber in all points along the polymer. However, it is not a pure sinusoidal function because of the jump conditions at the mechanical discontinuity points $x_1$ and $x_2$, cf.~Eqs.~\eqref{eq:jumps_1}-\eqref{eq:jumps_2}. Thus, in terms of the variable $\tilde x$ the eigenmodes are those of a uniform system with some jump conditions on the derivative. This is analogous, though not strictly equivalent, to the problem of a vibrating uniform string of length $\tilde L\!\equiv\! L \big(\phi\cdot\Delta+(1-\phi)\cdot1\big)$, with massive beads attached at the discontinuity points. For the latter, it is quite intuitive that $q_n\!\simeq\! n\pi/\tilde L$, which is the result of Eq.~\eqref{eq:slopes}.

Equations~\eqref{eq:DOS}-\eqref{eq:slopes}, together with the relation between $\lambda_q$ and $q_n$, provide an analytic description of the spectrum of eigenvalues in the framework of the continuum theory, which is the major result of this section. In the next section, the corresponding discrete problem is solved.

\section{Eigenmode analysis: Discrete theory}
\label{sec:DT}

In the preceding section the continuum eigenmode problem was formulated and solved. Here the same problem is addressed within the corresponding discrete theory, in order to highlight the similarities and the discrepancies between the two approaches. Our goal then is to find the eigenmodes $\vec w_q$, and their associated eigenvalues $\lambda_q$, that satisfy $\bm H \vec w_q\!=\!\lambda_q \vec w_q$. As before, we assume the eigenmodes to be sinusoidal with different wavelengths in the different regions. That is, we write the discrete analog of Eq.~\eqref{eq:eigenmodes}, where the $k$-th component $\vec w_q$ is given by
\begin{equation}
 \{w_q\}_k \sim \begin{cases}e^{i q k a} & \mbox{in the stiff regions}\\ e^{i \tilde q k a} & \mbox{in the soft regions}\end{cases} \ .
\end{equation}
For a homogeneous chain it is well known~\cite{AshcroftMermin}, and easily verified, that this results in a sinusoidal dispersion relation,
\begin{equation}
\label{eq:discrete_dispersion}
\begin{aligned}
\lambda(q)&=\kappa \left[2 \sin\left(\frac{qa}{2}\right)\right]^2 & \qquad\mbox{ for } \bm H\1\ , \\
\lambda(q)&=\kappa \left[2 \sin\left(\frac{qa}{2}\right)\right]^4 & \qquad\mbox{ for } \bm H\2\ .
\end{aligned}
\end{equation}
The allowed wavenumbers for homogeneous systems with the chosen boundary conditions are
\begin{equation}
 q_j a=\pi\frac{j-\frac{1}{2}}{N+\frac{1}{2}}\ , \qquad\qquad j=1,...,N\ .
 \label{eq:homq}
\end{equation}

Since the eigenvalue equation $H_{ij}w_j=\lambda(q)w_i$ must be satisfied with the same eigenvalue at all points, the relation between $q$ and $\tilde q$ (i.e.~the discrete analog of Eqs.~\eqref{eq:dispersion_1}-\eqref{eq:dispersion_2}) reads $\lambda(q)=\lambda(\tilde q)$. This implies
\begin{equation}
 \label{eq:dispersion_d}
\Delta\sin\left(\frac{qa}{2}\right)\!=\!\sin\left(\!\frac{\tilde{q}a}{2}\!\right)
\Rightarrow
\tilde{q}\!=\!\frac{2}{a}\sin^{-1}\!\left[\Delta\sin\left(\!\frac{qa}{2}\!\right)\right],
\end{equation}
valid for both $\bm H\1$ and $\bm H\2$. This identifies with Eqs.~\eqref{eq:dispersion_1}-\eqref{eq:dispersion_2} to leading order in $qa$, but differs substantially for $qa$ of order unity. Specifically, the sinusoidal functions can give rise to complex wavenumbers at high $qa$, that is, to (partially) evanescent eigenmodes. It is important to stress that this is a fundamental difference between the discrete and the continuum theories and that the discrete evanescent eigenmodes \textbf{do not have} a continuum counterpart. Physically, this happens because $\tilde q\!>\!q$ and therefore it might happen that at high $q$ the wavelength in the hard region is larger than the monomeric size $a$ (and thus is allowed), while in the wavelength in the soft region is shorter than $a$, and will thus be evanescent. One can see this explicitly by thinking of Eq.~\eqref{eq:dispersion_d} as an implicit function defining $\tilde q$ in terms of $q$. As $q$ grows, $\tilde q$ grows faster but this can only happen before the left-hand-side of Eq.~\eqref{eq:dispersion_d} reaches unity. For higher values of $q$ there exists no real solution for $\tilde q$. The transition occurs exactly when $\tilde qa\!=\!\pi$, i.e.~when the wavelength in the soft region is comparable to the monomeric size.

\begin{figure}
 \centering
 \includegraphics[width=\columnwidth]{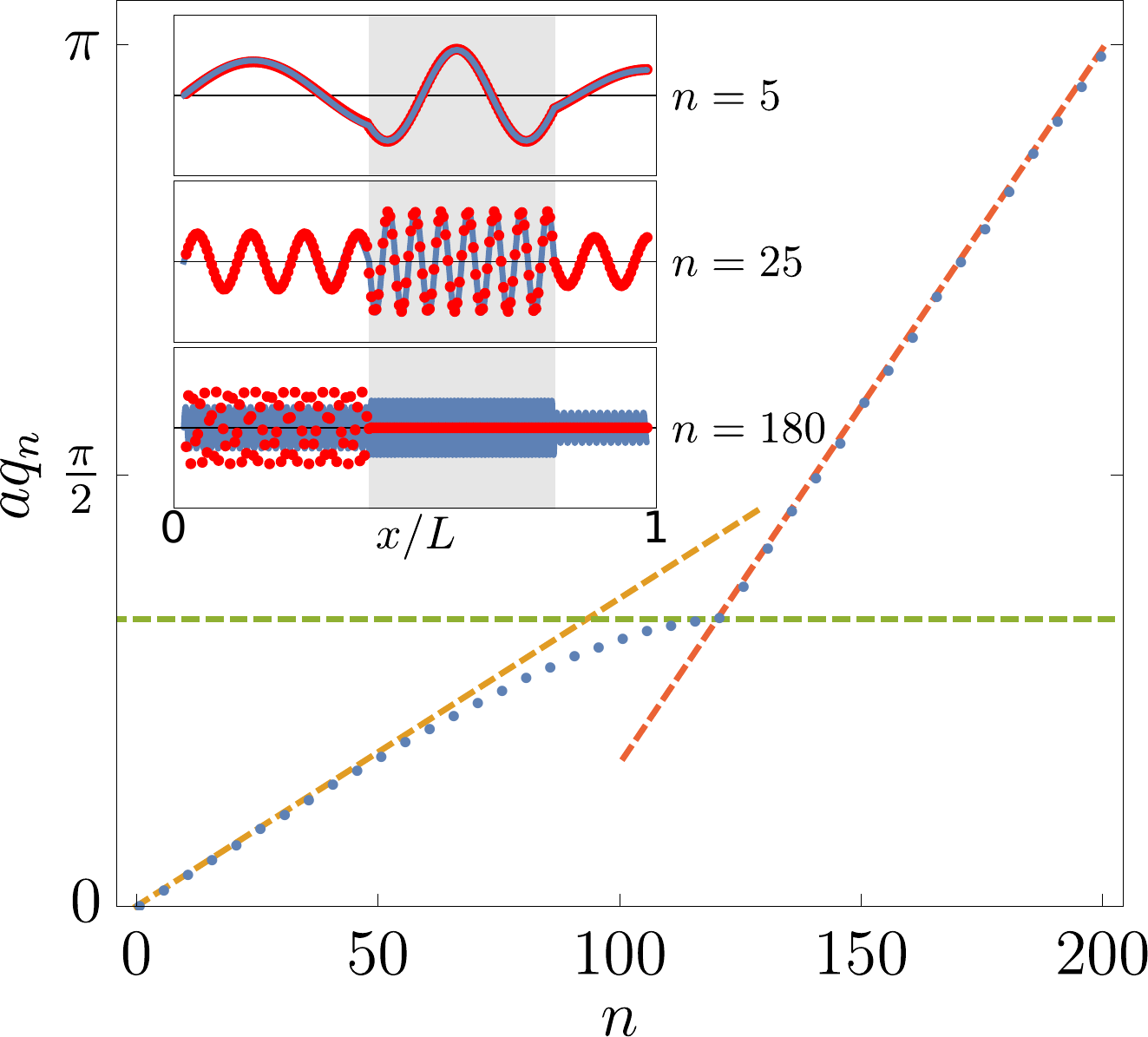}
 \caption{Spectrum of the discrete operator $\bm H\1$. Main panel: the $n$-th wavevector $q_n$ as a function of the ordinal number $n$. $q_n$ is obtained from the numerically calculated $\lambda_n$ by means of Eq.~\eqref{eq:discrete_dispersion}. For clarity, only every 5th value is plotted. The orange line shows the continuum prediction of Eqs.~\eqref{eq:DOS}-\eqref{eq:slopes}. The slope of the dashed red line corresponds to a homogeneous chain of length $(1-\phi)N$ (see text). The green line shows the value $\Delta\sin\!\left(\frac{qa}{2}\right)\!=\!1$, above which no real solution for $\tilde q$ exists, cf.~Eq.~\eqref{eq:dispersion_d}. Insets: A few selected eigenmodes (continuum theory in blue, discrete theory in red). It is seen that at low $q$ the agreement between the continuum and discrete theories is perfect, and that at higher $q$ discrepancies emerge. The 180th eigenmode is evanescent in the softer region, i.e.~$\tilde q$ is complex. The eigenvalues of $\bm H\2$ are almost identical to that of $\bm H\1$, but the eigenmodes differ (not shown).}
 \label{fig:dmodes}
\end{figure}

The existence of these evanescent high-$q$ modes is numerically verified, as shown in Fig.~\ref{fig:dmodes} along with the full spectrum. It is seen that the spectrum consists of two parts separated by a sharp boundary. This boundary corresponds exactly to the division between evanescent and non-evanescent modes and it occurs exactly at $qa\!=\!2\sin^{-1}\!\left(\Delta^{-1}\right)$, as predicted by Eq.~\eqref{eq:dispersion_d}. In fact, for the evanescent modes $q_n$ is linear in $n$, with a slope that identifies with that of a homogeneous chain, cf.~Eq.~\eqref{eq:homq}, when $N$ is replaced by the number of sites in the hard region, $N(1-\phi)$. This is demonstrated in Fig.~\ref{fig:dmodes}. With this, the eigenmode analysis in the framework of both the continuum and discrete theories is completed. Next, the statistical thermodynamic implications of the obtained results are explored.

\section{Statistical thermodynamics}
\label{sec:stat}

The statistical theory of fluctuating polymers has been intensively studied in the literature in various contexts~\cite{deGennes1979, Safran1994, Rubinstein2003}. Our goal here, following the analysis of the previous sections, is to understand the effect of spatial inhomogeneity on these fluctuations~\cite{Helfand1975, Su2010} and to elucidate the differences between the continuum and the discrete approaches to the problem. Specifically, we will address the dependence of thermodynamic quantities (mainly the free-energy) on the properties of the inclusion (i.e.~$\phi$ and $\Delta$) and on external geometric constraints, along with the associated fluctuation-induced forces.

A crucial player in theories of Gaussian thermal fluctuations is the eigenmode spectrum of the relevant dynamical operator. These spectra were analytically calculated in the preceding sections for both the continuum dynamical operator and its discrete counterpart. These calculations fully take into account the internal spatial inhomogeneity of the polymer, quantified by the normalized size $\phi$ and strength $\Delta$ of the inclusion. In addition, in order to account for prototypical external constraints, we focus on extensional fluctuations (i.e.~those governed by $U\1$), which are constrained by a rigid wall. Specifically, the relative elongation (strain) of the polymer is restricted to be smaller than $\varepsilon$, or equivalently, that its length is bounded to be smaller than $L(1+\varepsilon)$. In the limit $\varepsilon\!\to\!\infty$ the fluctuations are unconstrained, while otherwise the available configurations are constrained, which should be explicitly taken into account in thermal averages. In particular, if the field $w(x)$ is rendered dimensionless by measuring lengths in terms of $L$, the constraint is expressed mathematically by imposing $w_N\!<\!\varepsilon$ in the discrete description and $w(L)\!<\!\varepsilon$ in the continuous one. Since the results are qualitatively similar for both operators $U\1$ and $U\2$, we perform this analysis only for $U\1$, as stated above.

The main thermodynamic quantity of interest, from which all statistical thermodynamic properties follow, is the partition function ${\cal Z}$.
The parameters $\phi$, $\Delta$ and $\varepsilon$ affect ${\cal Z}$ in two distinct ways: The internal constraints, i.e. the properties of the inclusion $\phi$ and $\Delta$, affect the dynamical operator (and thus its spectrum) directly, while the external constraint $\varepsilon$ enters by restricting the allowed configurations over which the thermal average is performed. Explicitly, the partition function ${\cal Z}$ is given by the functional integral
\begin{equation}
{\cal Z}=\int Dw \,\exp\Big[-\beta \braket{w|\D(\phi,\Delta)|w}\Big] \Theta(\varepsilon-w(L))\ ,
 \label{eq:Z}
\end{equation}
where $\beta\!\equiv\!\left(k_B T\right)^{-1}$, $k_B$ is Boltzman's constant, $\Theta$ is Heaviside's step function and $\D$ is the dynamic operator, i.e.~either $\L\1$ or $\bm H\1$ (in the discrete calculation $w(L)$ should be replaced by $w_N$). For quadratic energy functionals, which is the subject of the present discussion, the partition function ${\cal Z}$ can be explicitly calculated in terms of Gaussian integrals. The calculation itself is rather straightforward, yet laborious. The details are given in the supplementary material~\cite{SM} and here we only discuss the final result in which the free-energy is expressed as
\begin{align}
 F&\equiv -k_BT\log{\cal Z}=F\uc(\phi,\Delta)+F_\varepsilon(\varepsilon,\phi,\Delta)\ ,
\label{eq:F}
\end{align}
where $F\uc$ is the free-energy of the unconstrained chain (i.e.~$F$ for $\varepsilon\!\to\!\infty$) and $F_\varepsilon$ is the contribution associated with the external constraint $\varepsilon$. Below we study each of these contributions separately.

\subsection{Unconstrained free-energy}

The unconstrained free-energy can be expressed in terms of the eigenvalues as~\cite{SM}
\begin{equation}
 F\uc\!=\!\frac{k_B T}{2}\log\left[\frac{\det\D}{(k_BT)^N}\right]
 \!=\!\frac{k_B T}{2}\sum_q \log\left(\frac{\lambda_q}{k_BT}\right)\ .
 \label{eq:free_energy}
\end{equation}
Equipped with an approximate expression for the eigenvalues of $\L\1$ and an analytic expression for $\det \bm H\1$, the above formula can be evaluated explicitly~\cite{SM}. The result, after taking the large-$N$ limit, reads
\begin{align}
\label{eq:F_uc}
 F\uc\DT&\!=\!N k_B T\left[\frac{1}{2}\log\left(\frac{\beta \kappa_h}{L/N}\right)\!-\!\phi \log\Delta \right]\ ,\\
 F\uc \SFT&\!=\!N k_B T~\times\\
  &\!\!\left[\frac{1}{2}\log\left(\frac{\beta \kappa_h}{L/N}\right)\!-\!\log \left(\phi\Delta+1-\phi\right) \!+\! \log\left(\frac{\pi\sqrt{N}}{e}\right)\right].\nonumber
\end{align}
Here and in what follows the superscript DT stands for ``Discrete Theory'', i.e.~results pertaining to $\bm H\1$, and CT for ``Continuum Theory'', i.e.~results pertaining to $\L\1$.

\begin{figure}[h]
 \centering
 \includegraphics[width=\columnwidth]{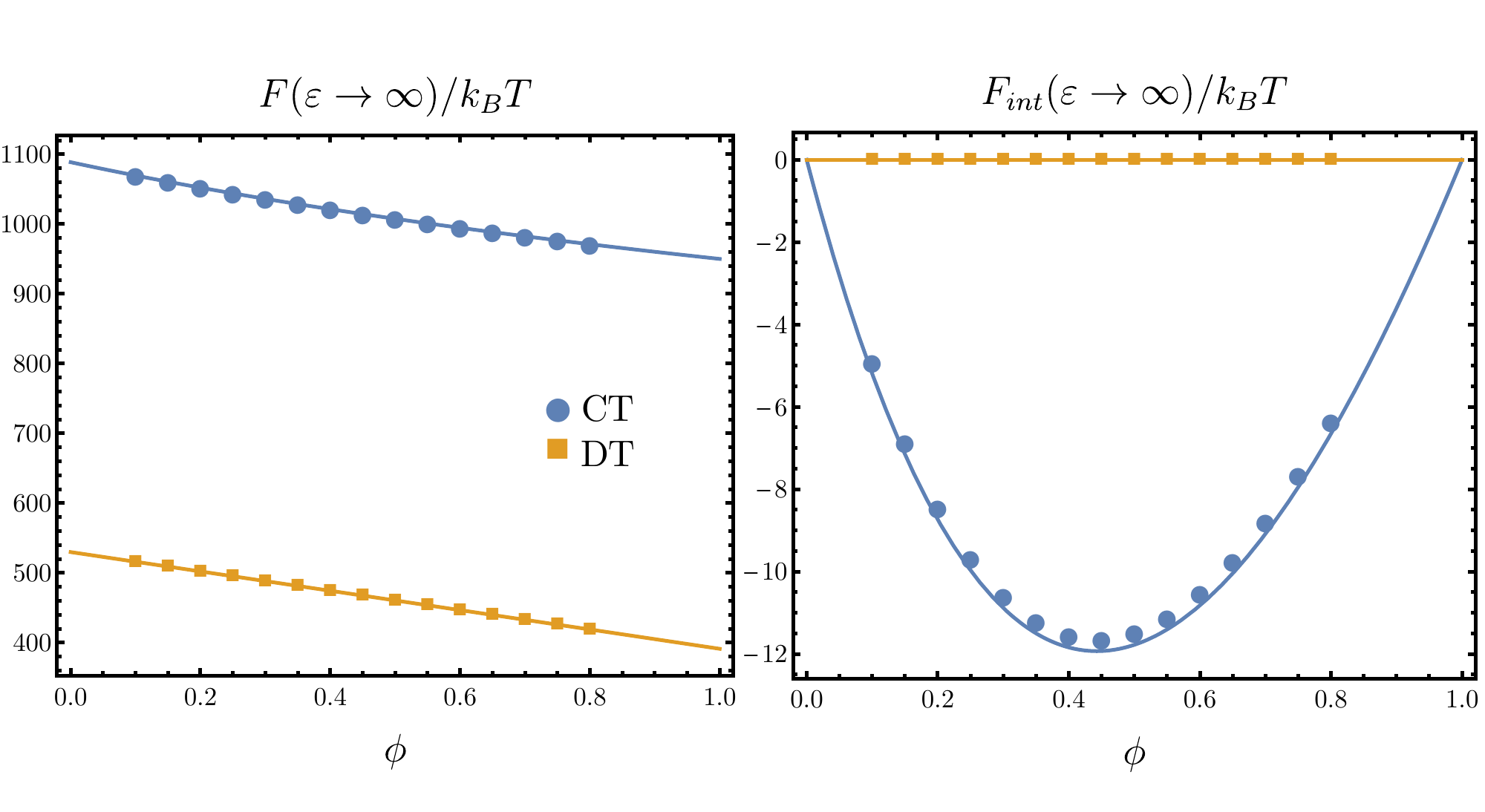}
 \caption{The total free-energy (left) and the interaction free-energy (right) as a function of $\phi$ for $\varepsilon\!\to\!\infty$. The solid lines show the analytic predictions of Eqs.~\eqref{eq:F}-\eqref{eq:fint} and the data points correspond to direct numerical calculations. The small negative deviation of the continuum theory prediction for $F\sint$ from the numerical results emerges from using Stirling's approximation, which can be eliminated by taking higher order corrections (not shown).}
 \label{fig:F}
\end{figure}

To gain more insight into the structure and physical content of Eqs.~\eqref{eq:F_uc}, we rewrite the unconstrained free-energy as the sum of the free-energies of the homogeneous segments and an interaction energy. That is, we write
\begin{equation}
\label{eq:split}
 F\uc=N\Big(\phi f(\kappa_s) +(1-\phi)f(\kappa_h)\Big)+F\sint\ ,
\end{equation}
for both theories, where $f(\kappa)$ is the specific (per monomer) free-energy of a homogeneous polymer with modulus $\kappa$, and $F\sint$ is the interaction energy between the soft and the hard segments. This form of writing is common in the context of calculating Casimir-like fluctuation-induced forces between inclusions~\cite{Goulian1993,Golestanian1996,Kardar1999}, to be discussed below. In this representation, we need to calculate the homogeneous polymer free-energies in the two theories, which take the form~\cite{SM}
\begin{equation}
\label{eq:fh}
\begin{split}
 f\DT(\kappa)&=\frac{1}{2}k_BT\log\left(\frac{\beta\kappa}{a}\right)\ ,\\
 f\SFT(\kappa)&=\frac{1}{2}k_BT\log\left(\frac{\beta\kappa}{a}\right)+k_BT\log\left(\frac{\pi\sqrt N}{e}\right)\ .
\end{split}
\end{equation}
We note that the two theories agree quantitatively on the specific free-energy, up to a logarithmic factor in $N$. The latter actually implies that the free-energy in the continuum theory is not strictly extensive, an issue that pertains already to the continuum theory of homogeneous systems and is not discussed here. Equations~\eqref{eq:F_uc}-\eqref{eq:fh} indicate that the interaction energy in the two cases reads
\begin{equation}
\label{eq:fint}
 F\DT\sint\!=\!0\ , \qquad F\SFT\sint\!=\!N k_BT\log \left(\frac{\Delta ^{\phi}}{\phi \Delta +(1-\phi )}\right)\ ,
\end{equation}
revealing fundamental differences between the two theories. This non-trivial result means that the discrete theory predicts the free-energy of an inhomogeneous polymer to be simply the sum of the free-energies of the soft and hard regions without any interaction. In fact, this holds for an arbitrary choice of $\kappa_i$, not necessarily the hard-soft-hard configuration described here~\cite{SM}. In contrast, the continuum theory predicts a non-trivial dependence of the free-energy on the inclusion parameters. The analytic results in Eqs.~\eqref{eq:F}-\eqref{eq:fint} are all corroborated against explicit numerical calculations, as shown in Fig.~\ref{fig:F}.

This discrepancy in the free-energy can be manifested in measurable quantities, such as the configurational contribution to the fluctuation-induced force $\partial_\phi F$.
Physically, this force corresponds --- e.g.~in the case of cofilin-mediated softening of actin filaments, where local softening of the actin polymer is induced by the adsorption of cofilin molecules from the solvent~\cite{McCullough2008} --- to the fluctuation-induced contribution to an adsorption force. The latter also includes other contributions, e.g.~a binding energy, the change in the solvent mixing entropy and the entropy associated with placing the inclusion at different locations along the polymer, which are of no interest in the present context.

As we focused here on the unconstrained free-energy $F\uc$, we calculate first $\chi_\phi\equiv\pa_\phi F\uc$, which takes the form
\begin{equation}
 \label{eq:chi}
 \begin{split}
 \chi_\phi\DT&=-N k_B T\log \Delta, \\
 \chi_\phi\SFT&=-N k_B T\frac{\Delta -1}{(\Delta -1) \phi +1} \ .
 \end{split}
\end{equation}
Later on we will show that the contribution of $F_\varepsilon$ to this force, $\pa_\phi F_\varepsilon$, is the same for both theories. $\chi_\phi\DT$ and $\chi_\phi\SFT$ of Eq.~\eqref{eq:chi} agree only in the limit of very small mechanical contrast, $\Delta\!\to\!1$, but otherwise significantly differ, highlighting a stark discrepancy between the continuum and discrete theories. This discrepancy will be extensively discussed in Sec.~\ref{sec:comp}. Before that, we study the free-energy associated with the external constraint $\varepsilon$ and see whether similar discrepancies persist there too.

\subsection{External-constraint-related free-energy}

The contribution of the external constraint to the free-energy, $F_\varepsilon$, can be explicitly calculated~\cite{SM} and takes the form
\begin{equation}
\label{eq:F_ve}
\begin{split}
 F_\varepsilon&=k_BT\log\left[1+\erf\left(\varepsilon\sqrt{\frac{\beta \kappa\eff(\phi,\Delta)}{2L}}\right)\right]\ .
\end{split}
\end{equation}
This expression for $F_\varepsilon$ is valid for both the continuum and the discrete theories, where the effective modulus that depends on the inclusion parameters $\kappa\eff(\phi,\Delta)$ takes the form 
\begin{equation}
\label{eq:kappa_eff}
 \begin{aligned}
  \kappa\DT\eff(\phi,\Delta)&=\left(\frac{\phi}{\kappa_s}+\frac{1-\phi}{\kappa_h}\right)^{-1}
  =\frac{\kappa_h}{\phi \Delta^2+(1-\phi)}\ ,\\
  \kappa\SFT\eff(\phi,\Delta)&=\kappa_h\left[\sum_q \left(\frac{u_q(L)}{q L}\right)^2\right]^{-1}\ ,
 \end{aligned}
\end{equation}
in the two theories. $\erf(\cdot)$ in Eq.~\eqref{eq:F_ve} is the error function.

\begin{figure*}
\centering
  \includegraphics[width=\textwidth]{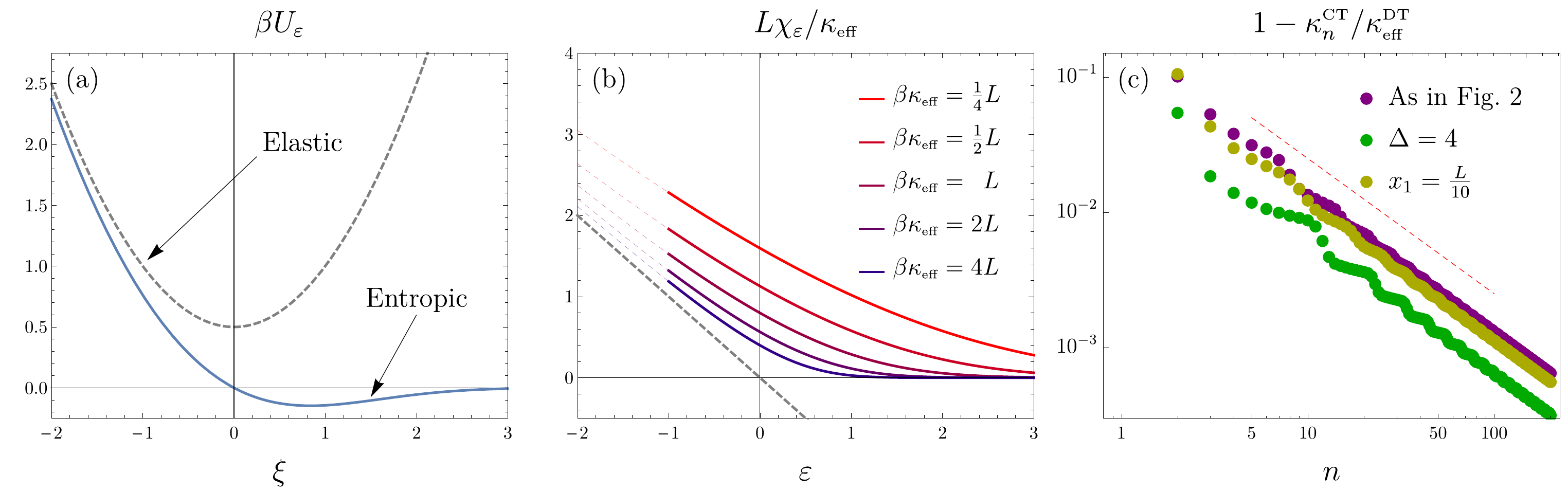}
 \caption{(a) The internal energy $U_\varepsilon$ related to the external constraint $\varepsilon$ (solid line), cf.~Eq.~\eqref{eq:Uvarepsilon}, and the elastic energy with effective modulus $\kappa\eff$ (dashed line), cf.~Eq.~\eqref{eq:Uvarepsilon1}, as a function of the dimensionless coordinate $\xi$. The elastic energy is plotted also for $\xi\!>\!0$, corresponding to extension, though no elastic behavior is physically observed in this regime as the confining wall is not attached to the polymer. (b) The thermodynamic force $\chi_\varepsilon$ as a function of $\varepsilon$ for varying values of the dimensionless combination $\beta\kappa\eff/L$. While results for $\varepsilon\!<\!-1$ are outside of the physically accessible range and hence are plotted in thin dashed lines, they provide good approximations to the behavior of the free-energy in the physical range $\varepsilon\gtrsim-1$. (c) The convergence of the continuum theory prediction for $\kappa\eff$ towards the discrete theory prediction of Eq.~\eqref{eq:kappa_eff}. The plot shows the relative deviation $1-\kappa_n\SFT/\kappa\eff\DT$, where the partial sum $\kappa_n\SFT$ is defined in Eq.~\eqref{eq:partialsum}, as a function of the number of summed eigenmodes $n$. The purple data correspond to the same parameter values used throughout the paper, e.g.~Fig.~\ref{fig:modes}, and other colors correspond to one parameter being changed each time, as stated in the legend. It is observed that a convergence to within 1\% is achieved after summing over the first $\sim\!10$ modes for all parameters tested.}
 \label{fig:varepsilon}
\end{figure*}

Since the two theories predict the same functional form for $F_\varepsilon$, differences between them can emerge only due to possible differences between $\kappa\eff\DT$ and $\kappa\eff\SFT$. We thus need to compare these two. $\kappa\eff\DT$ in Eq.~\eqref{eq:kappa_eff} is exactly the effective macroscopic $\kappa$ of a chain of microscopic springs connected in series. To better understand $\kappa\eff\SFT$ and its relation to $\kappa\eff\DT$, we define $\kappa\SFT_n$ as the partial sum over eignemodes
\begin{equation}
\label{eq:partialsum}
\kappa\SFT_n=\kappa_h\left[\sum_{i=1}^n \left(\frac{u_{q_i}(L)}{q_i L}\right)^2\right]^{-1}\ .
\end{equation}
In this way we can quantify the contribution of eigenmodes of increasing wavenumber to $\kappa\eff\SFT$.
In Fig.~\ref{fig:varepsilon}c we plot the deviation of $\kappa_n\SFT/\kappa\eff\DT$ from unity as a function of the number of modes $n$. It is observed that $\kappa_n\SFT$ converges to the discrete theory prediction $\kappa\DT\eff$ after summation over a sub-extensive number of modes. That is, the two theories essentially predict the same effective modulus $\kappa\eff$ and consequently the same $F_\varepsilon$. This agreement, contrasted with the discrepancy in the two predictions for $F\uc$, will be discussed in Sec.~\ref{sec:comp}.

Before concluding this subsection, let us briefly comment on the structure of $F_\varepsilon$, which has a neat physical interpretation. Let us consider the internal energy $U_\varepsilon\!=\!-\pa_{\beta}(\beta F_\varepsilon)$ (which is the same for both the continuum and the discrete approaches), which reads
\begin{align}
\label{eq:Uvarepsilon}
\beta U_\varepsilon(\xi)=-\frac{e^{-\xi ^2} \xi }{\sqrt{\pi}\left[1+\erf(\xi)\right]}\ ,
\end{align}
where the notation $\xi\!\equiv\!\varepsilon\sqrt{\frac{\beta\kappa\eff}{2L}}$ was introduced. We note that the internal energy associated with the unconstrained free-energy, $\pa_{\beta}(\beta F\uc)$, trivially equals $\tfrac{1}{2}N k_BT$ according to the equipartition theorem. Consequently, $U_\varepsilon$ in fact measures the deviation of the internal energy from the background thermal energy predicted by equipartition.

In the limit of large $\xi$, $U_\varepsilon(\xi)$ vanishes, as expected (i.e. the polymer is essentially unconstrained). In the limit of large negative $\xi$ (note, though, that $\varepsilon$ is physically bounded from below by $-1$), we have
\begin{align}
\label{eq:Uvarepsilon1}
U_\varepsilon(\xi\to-\infty)\simeq k_B T\left(\xi^2+\frac{1}{2}\right)=\frac{\kappa\eff}{2L}\varepsilon^2+\frac{k_B T}{2}\ .
\end{align}
In this limit, the polymer is under compression and responds predominantly elastically, i.e. its internal energy varies as $\varepsilon^2$ with a prefactor proportional to the effective modulus $\kappa\eff$. Note that the ordinary compression-extension elastic symmetry, i.e. symmetry under $\varepsilon\!\to\!-\varepsilon$ ($\xi\!\to\!-\xi$), is broken here since the confining wall is not attached to the polymer. All of the properties of $U_\varepsilon(\xi)$ are shown in Fig.~\ref{fig:varepsilon}a. When $\xi$ is not very negative $F_\varepsilon$ is entropic in nature and vanishes for $T\!\to\!0$ (recall that the persistence length of a homogeneous polymer is $\beta\kappa$, hence the factor $\frac{\beta\kappa\eff}{2L}$ can be interpreted as the number of times the effective persistence length enters in the size of the polymer).

The thermodynamic force related to the external constraint, $\chi_\varepsilon\equiv\pa_\varepsilon F$, is a measurable physical quantity (e.g.~the pressure on a confining wall) that can also be calculated. It is plotted in Fig.~\ref{fig:varepsilon}b, where it is seen that for negative values of $\varepsilon$ near $-1$ it is linear and its origin is predominantly elastic, as expected from the preceding discussion, while it decays to zero when $\varepsilon\!\to\!\infty$. For intermediate positive values it is a fluctuation-induced force and the transition between the elastic and fluctuation-induced regimes is not sharp, but is rather smoothed by the temperature. Clearly, for $T\!\to\!0$ the force is strictly linear at $\varepsilon\!<\!0$ and strictly vanishes for $\varepsilon\!>\!0$.

Next, we turn to discuss the relations between the continuum and discrete theories in light of the results obtained up to now.

\section{Validity of the continuum theory}
\label{sec:comp}

In the previous section we saw that various statistical thermodynamic properties of inhomogeneous polymers reveal significant differences between the continuum and discrete theories. That is, the interaction free-energy between the soft and hard segments in Eq.~\eqref{eq:fint} and the fluctuation-induced adsorption force in Eq.~\eqref{eq:chi} feature qualitative discrepancies between the continuum and discrete theories, except for the small contrast limit $\Delta\!\to\!1$, where $F\SFT\sint\!\approx\!F\DT\sint\!=\!0$ and $\chi_\phi\SFT\!\approx\!\chi_\phi\DT\!\sim\!\Delta\!-\!1$. In particular, $F\sint\DT\!=\!0$ identically, while its continuum counterpart is a non-trivial function of $\phi$ and $\Delta$, cf.~Eq.~\eqref{eq:fint}. A corollary is that $\chi_\phi\SFT$ depends on $\phi$, while $\chi_\phi\DT$ is independent of it, cf.~Eq.~\eqref{eq:chi}. We stress that these discrepancies are not mitigated when the discretization length is taken to zero, when a different ultra-violet cutoff is used or when the variation of $\kappa(x)$ is smoothed out~\cite{SM}. What can one make of these discrepancies?

Obviously, the continuum analysis of the eigenmodes and eigenvalues of the inhomogeneous polymer, which is the basis for any statistical thermodynamic calculation in the Gaussian approximation, is strictly valid only for small wavenumbers $qa\!\ll\!1$. This is true in general and has been analytically demonstrated in sections \ref{sec:CT} and \ref{sec:DT}, revealing qualitative differences in the eigenmodes and eigenvalues spectra of the continuum and discrete operators at large wavenumbers. This by itself does not invalidate the continuum approach. The pertinent question then is whether a given physical observable is dominated by small wavenumbers, in which case the continuum approximation is valid.

The results of section \ref{sec:stat} indicate that this is not the case. Beyond the directly observed differences between the continuum and discrete results themselves, this can be inferred from the continuum result alone. Let us go back to Eq.~\eqref{eq:split}; the first two contributions to the (unconstrained) free-energy on the right-hand-side are ``bulk'' contributions, i.e. terms that scale with the total size of the soft segment $\phi N$ and the two hard segments $(1-\phi)N$. Since $N\!\sim\!L/a\!\sim\!L q_{max}$, where $q_{max}$ is the UV-cutoff, these contributions depend explicitly on the large wavenumbers and in general are not expected to be correctly described by the continuum theory (we note again that the fact that the continuum ``bulk'' free-energy is not even strictly extensive in our case, cf.~the second equation in~\eqref{eq:fh}, is not discussed here). The important point is that in the thermodynamic limit, where $q_{max}\!\to\!\infty$, these ``bulk'' contributions diverge and are commonly eliminated in standard calculations~\cite{Kardar1999, Simpson2015}.

We are then left with the last term on the right-hand-side of Eq.~\eqref{eq:split}, $F\SFT\sint$, the interaction free-energy between soft and hard segments. The result in Eq.~\eqref{eq:fint} shows that the interaction free-energy also scales with $N\!\sim\!q_{max}$ and hence depends explicitly on the UV-cutoff, marking the breakdown of the continuum theory in this case. Consequently, the continuum result for the fluctuation-induced adsorption force, $\chi_\phi\SFT$ in Eq.~\eqref{eq:chi}, scales with the system size $N$ and is therefore not dominated by small wavenumbers. This should be contrasted with Casimir-like fluctuation-induced forces in which the interaction energy depends on a geometric degree of freedom, e.g. the separation between two plates, but is independent of $q_{max}$~\cite{Casimir1948, Simpson2015}. In this case, after the divergent ``bulk'' contributions are removed, a continuum-level fluctuation-induced force is identified by taking the derivative of the interaction free-energy with respect to the geometric degree of freedom. It is important to note that a physically realistic fluctuation-induced adsorption force does exist in our problem and is given by the discrete theory result $\chi_\phi\DT$ in Eq.~\eqref{eq:chi}.

The continuum analysis presented above bears some similarity to the Debye model of the specific heat of homogeneous systems. There, similarly to the main panel of Fig.~\ref{fig:dmodes}, a continuum-level linear spectrum of wavenumbers replaces the nonlinear spectrum of the discrete theory (both agree of course for $qa\!\ll\!1$), keeping the total number of eigenmodes the same. When coupled to the Bose-Einstein statistics for the occupation number the heat capacity features the famous $T^3$ behavior at low $T$. There are two major differences between Debye's analysis and ours; first, our analysis was strictly classical, not taking into account quantum effects such as those incorporated into the Bose-Einstein distribution. This makes a difference because the latter provides a physical UV cutoff that at low $T$ assigns negligible weight to the high-$q$ modes for which the continuum theory is invalid. Second, as we explicitly demonstrated, spatial inhomogeneity gives rise to differences between the continuum and discrete eigenmodes/eignevalues which are not encountered in spatially homogeneous systems.

In contrast to $\chi_\phi$, the fluctuation-induced force associated with an external constraint --- $\chi_\varepsilon$ derived from $F_\varepsilon$ of Eq.~\eqref{eq:F_ve} --- does not depend on $q_{max}$ and the continuum and discrete predictions coincide. That is, this fluctuation-induced force is properly described by the continuum theory. The reason for this is that in this case the relevant fluctuations are shape fluctuations, which are dominated by small wavenumbers (since the amplitude of the eigenmodes decays with increasing $q$). Mathematically speaking, this property is encapsulated in the fact that $\kappa\SFT\eff$ of Eq.~\eqref{eq:kappa_eff}, which is expressed as a sum over wavenumbers, converges to $\kappa\DT\eff$ of Eq.~\eqref{eq:kappa_eff} after summing over the first few smallest wavenumbers, as shown in Fig.~\ref{fig:varepsilon}.

\section{Concluding remarks}

In this paper we studied the mechanics and statistical thermodynamics of semiflexible inhomogeneous polymers. We focused on inhomogeneity in the form of a soft inclusion embedded inside a stiffer/harder polymer, and considered torsional, extensional and bending Gaussian fluctuations. Analytic results for the eigenmodes and eigenvalues spectra of both the continuum and the corresponding discrete dynamical operators were derived. The analysis revealed qualitative differences between the continuum and discrete spectra. Most notably, it was shown that above a certain wavenumber, the discrete spectrum of wavenumbers $q_n$ changes qualitatively and the discrete modes become evanescent inside the soft inclusion, having no continuum counterparts.

Based on the eigenmodes and eigenvalues analysis, we derived explicit expressions for two types of fluctuation-induced forces in the framework of both the continuum and discrete theories. One fluctuation-induced force is associated with variations of the properties of the inclusion, i.e. its size and strength. This entropic force describes, for example, the fluctuation-induced contribution to the adsorption of molecules that give rise to the soft inclusion. Another fluctuation-induced force is associated with an external geometric constraint, i.e. a confining wall with variable position. This entropic force describes the pressure applied by the fluctuating polymer on the wall.

It was shown that the first fluctuation-induced force is dominated by contributions from modes with large wavenumbers, where the continumm and discrete spectra significantly differ, and hence that the continuum theory breaks down. On the other hand, the second fluctuation-induced force was shown to be dominated by small wavenumber shape fluctuations and hence is properly described by the continuum theory. The results show that while the continuum theory of inhomogeneous polymers may be successful in some cases, it fails in others, and should be taken with some caution.

\textit{Acknowledgments} We are indebted to S.~Safran and O.~Farago for very useful discussions. We also thank E.~Brener, M. Kardar, N.~Gov and P.~Pincus for useful comments. E.~B.~acknowledges support from the Israel Science Foundation (Grant No.~712/12), the Harold Perlman Family Foundation and the William Z. and Eda Bess Novick Young Scientist Fund.


\onecolumngrid
\newpage
\begin{center}
	\textbf{\large Supplementary Material for:\\``Gaussian fluctuations of spatially inhomogeneous polymers''}
\end{center}
\twocolumngrid

\setcounter{equation}{0}
\setcounter{figure}{0}
\setcounter{section}{0}
\setcounter{table}{0}
\setcounter{page}{1}
\makeatletter
\renewcommand{\theequation}{S\arabic{equation}}
\renewcommand{\thefigure}{S\arabic{figure}}
\renewcommand{\thesection}{S-\Roman{section}}
\renewcommand*{\thepage}{S\arabic{page}}
\renewcommand{\bibnumfmt}[1]{[S#1]}
\renewcommand{\citenumfont}[1]{S#1}

This document is meant to provide additional technical details related to results reported on in the manuscript.

\section{Free energy in the continuum theory}
Consider a general one-dimensional system whose energy is treated to quadratic order. The continuum energy takes the form
\begin{align}
U\SFT(w(x))&\!=\!\frac{1}{2}\braket{w(x)|\L|w(x)}\!\equiv\!\frac{1}{2L}\int_{0}^{L} w(x)\, \L \,w(x)\, dx\ ,\nonumber
\end{align}
where $\L$ is a self-adjoint real differential operator. Our convention is that the eigenvalues of $\L$ are of energy dimensions, and thus $w(x)$ is dimensionless. We work in the eigenbasis of $\L$, which we denote by $w_{q_1}(x),\dots,w_{q_N}(x)$. These functions are orthonormal, i.e.
\begin{equation}
\begin{aligned}
&\frac{1}{L}\int_0^L w_q w_{q'} dx = \delta_{qq'} \ ,\quad \mbox{and }\\
&\braket{w_q|\L|w_{q'}}=\lambda_q \delta_{qq'}\ ,
\end{aligned}
\end{equation}
where $\lambda_q$ is the eigenvalue associated with $w_q$. The eigenmodes span the functional space and a general configuration $w(x)$ can be written as $w(x)=\sum_q a_q w_q(x)$ where $a_q\equiv \braket{w(x)|w_{q}}$. The energy is thus written as
$ U\SFT=  \frac{1}{2} \sum_q \lambda_q\,a_q^2$, and the partition function, defined in Eq.~\eqref{eq:Z} of the main text, reads
\begin{align}
&Z\SFT=\int D w \,  e^{-\beta \braket{w|\L|w}} \Theta\Big(\varepsilon-w(L)\Big) \label{eq:Z_disc_SM}\\
&=\int d^N a_q \,  \exp\left[-\beta\sum_q \tfrac{1}{2}\lambda_qa_q^2\right] \Theta\Big(\varepsilon-\sum_q a_q w_q(L)\Big)\ .
\nonumber
\end{align}
This is a multivariate Gaussian integral over a half-space. In Sec.~\ref{sec:gauss} of this file we derive a general formula for integrals of this type (Eq.~\eqref{eq:half_space}). Applying this formula to Eq.~\eqref{eq:Z_disc_SM} yields
\begin{align}
Z\SFT&=\frac{1}{2} \left(\frac{(2\pi)^N}{\beta^N \det \L}\right)^{1/2}\left(1+\erf\left[\varepsilon \sqrt{\frac{\beta \kappa_h}{\ell\SFT}}\right]\right)\\
\ell\SFT&\equiv 2 L\sum_q \left(\frac{w_q(L)}{q L}\right)^2\ . \label{eq:ell_SFT}
\end{align}
where $\erf(\cdot)$ is the standard error function and the relation $\lambda_q=\kappa_h L q^2$ was used. $\det\L$ is defined as $\prod_q\lambda_q$. The factor $\frac{1}{2}(2\pi)^{N/2}$ is of no physical importance and will be omitted in what follows.

Note that here we take into account exactly $N$ continuum modes, which is basically a choice of an ultraviolet cutoff on $q$. The results presented here do not depend qualitatively on the choice of the ultraviolet cutoff, as long as the number of modes scales with $N$, which is anyway a trivial requirement from any reasonable cutoff scheme.

The free energy is thus given by
\begin{equation}
\begin{split}
F\SFT&\equiv- k_B T \log Z\SFT=F\SFT\uc+F\SFT_\varepsilon \ ,\\
F\SFT\uc &\equiv \tfrac{1}{2}k_BT\log\left(\beta^N \det \L\right) \ ,\\
F\SFT_\varepsilon&\equiv -\tfrac{1}{2}k_BT\log\left(1+\erf\left[\varepsilon \sqrt{\frac{\beta \kappa_h}{\ell\SFT}}\right]\right)\ . 
\end{split}
\end{equation}

We now turn to calculate $\det\L$, which is done by explicit calculation of the eigenmodes. Since the wavenumbers are given approximately by Eqs.~\eqref{eq:DOS}-\eqref{eq:slopes} of the main text, the calculation of $F\SFT$ is straightforward. The eigenvalue associated with the wavenumber $q$ is $\lambda_q =L\,\kappa_h\, q^2$ and thus
\begin{align*}
\det\L&=\prod_{q}\lambda_q =
\prod_{n}\kappa_h L q_n^2=
\left(\frac{\kappa_h}{L}\right)^N C^{2N} (N!)^2 \ .
\end{align*}
This immediately leads to
\begin{align}
&F\SFT=\frac{1}{2}k_B T \log\left[\beta^N\det\L\right]=\\
&N k_B T\left[\tfrac{1}{2}\log\left(\frac{\beta \kappa_h}{L}\right) -\log\left(\frac{\phi\Delta+1-\phi}{\pi}\right) +\frac{1}{N}\log{N!} \right]\nonumber.
\end{align}
We now apply Stirling's approximation, which we write as $\log(N!)\approx N\log\left(\frac{N}{e}\right)$ and after some rearrangement we obtain
\begin{equation}
\begin{split}
F\SFT\approx\,
&N k_B T\Bigg[\tfrac{1}{2}\log\left(\frac{\beta \kappa_h}{L/N}\right) \\&-\log\left(\Delta\phi+1-\phi\right) +\log\left(\sqrt{N}\frac{\pi}{e}\right)\Bigg] \ .
\end{split}
\end{equation}
The free energy of a homogeneous polymer, $f\SFT$ is immediately obtained by setting $\phi=0$.

\section{Free energy in the discrete theory}
\label{sec:discrete_calculation}
Here we present the calculation of the free energy associated with $\bm H\1$ in the discrete formalism. We want to calculate the partition function
\begin{equation}
Z\DT=\int_{-\infty}^\infty d^N \bm w \,  e^{-\beta U\DT(\bm w)} \Theta(\varepsilon-w_N) \ ,
\end{equation}
with 
\begin{equation}
U\DT(\bm w)=\sum_{i=1}^N\frac{1}{2}\kappa_i\left(\frac{w_i-w_{i-1}}{\Delta x}\right)^2\Delta x\ .
\end{equation}
Unlike the continuum case described in the previous section, here the calculation can be performed without explicit reference to the eigenmodes. The trick is to use the non-orthogonal change of variables
\begin{align}
y_i&\equiv \sqrt{\frac{\kappa_i}{\Delta x}}(w_i-w_{i-1})\ ,
&
w_i&= \sum_{j=1}^i y_j\sqrt{\frac{\Delta x}{\kappa_j}}\ .
\end{align}
The Jacobian of this transformation is $\prod_i \sqrt{\frac{\kappa_i}{\Delta x}}=\sqrt{\det \bm H}$. With the new variables $y_i$ the energy takes the simple form $U=\frac{1}{2}||\bm y||^2$. The partition function is thus
\begin{align*}
Z\DT&=\sqrt{\frac{1}{\det\bm H}}\int_{-\infty}^\infty d^N\bm y \,  e^{-\frac{\beta}{2}|\bm y|^2}\Theta \left(\varepsilon-\sum_j y_j\sqrt{\frac{\Delta x}{\kappa_j}}\right)\ .
\end{align*}
This is a Gaussian integral over a half-space, for which we derive an explicit formula in Sec.~\ref{sec:gauss} of this file (Eq.~\eqref{eq:half_space}). The result is
\newcommand{\keff}{\kappa\eff}

\begin{align}
Z\DT&=\frac{1}{2}\sqrt{\frac{(2\pi)^N}{\beta^N \det\bm H}}\left(1+\erf\left[\varepsilon \sqrt{\frac{\beta \keff}{2 \Delta x}}\right]\right)\ ,
\end{align}
where we introduced the notation $\keff\equiv\left(\sum \kappa_i^{-1}\right)^{-1}$, i.e.~the effective spring constant of the chain.

Note that this expression holds for an arbitrary choice of $\kappa_i$, and also that it is invariant to permutations in the order of the $\kappa_i$'s (since $\det\bm H$ is). If we assume $\kappa(x)$ has the form described in the main text, i.e.~$N\phi$ springs have a spring constant of $\kappa_s$ and $N(1-\phi)$ have a spring constant of $\kappa_h$, we have
\begin{align}
&\keff=\left(\frac{N\phi}{\kappa_s}+\frac{N(1-\phi)}{\kappa_h}\right)^{-1}=\frac{\kappa_h/N}{\Delta^2\phi+(1-\phi)}\ ,\\
&\det\bm H=\prod_i \frac{\kappa_i}{\Delta x}=\left(\frac{\kappa_h}{\Delta x}\right)^N\Delta^{-2N\phi}\ .
\end{align}

Thus, the free energy is 
\begin{align}
\nonumber 	F\DT&\equiv-k_B T \log Z\DT=F\DT_i+F\DT_\varepsilon \ ,\\
\nonumber 	F\DT\uc &\equiv \tfrac{1}{2}k_BT\log\left(\beta^N \det \bm H\right)\\
\label{eq:FDT}	&=\tfrac{N}{2}k_BT\left[\log\left(\frac{\beta\kappa_h}{L/N}\right) -\phi\log\Delta\right]\ ,\\
\nonumber	F\SFT_\varepsilon&\equiv -\tfrac{1}{2}k_BT\log\left(1+\erf\left[\varepsilon \sqrt{\frac{\beta \kappa_h}{\ell\DT}}\right]\right)\ ,\\
\ell\DT&=2L\Big(\Delta^2\phi+(1-\phi)\Big)\ .
\label{eq:ell_DT}
\end{align}

\section{Smooth variation of $\kappa(x)$}
\label{sec:smooth}

\begin{figure}
	\centering
	\includegraphics[width=\columnwidth]{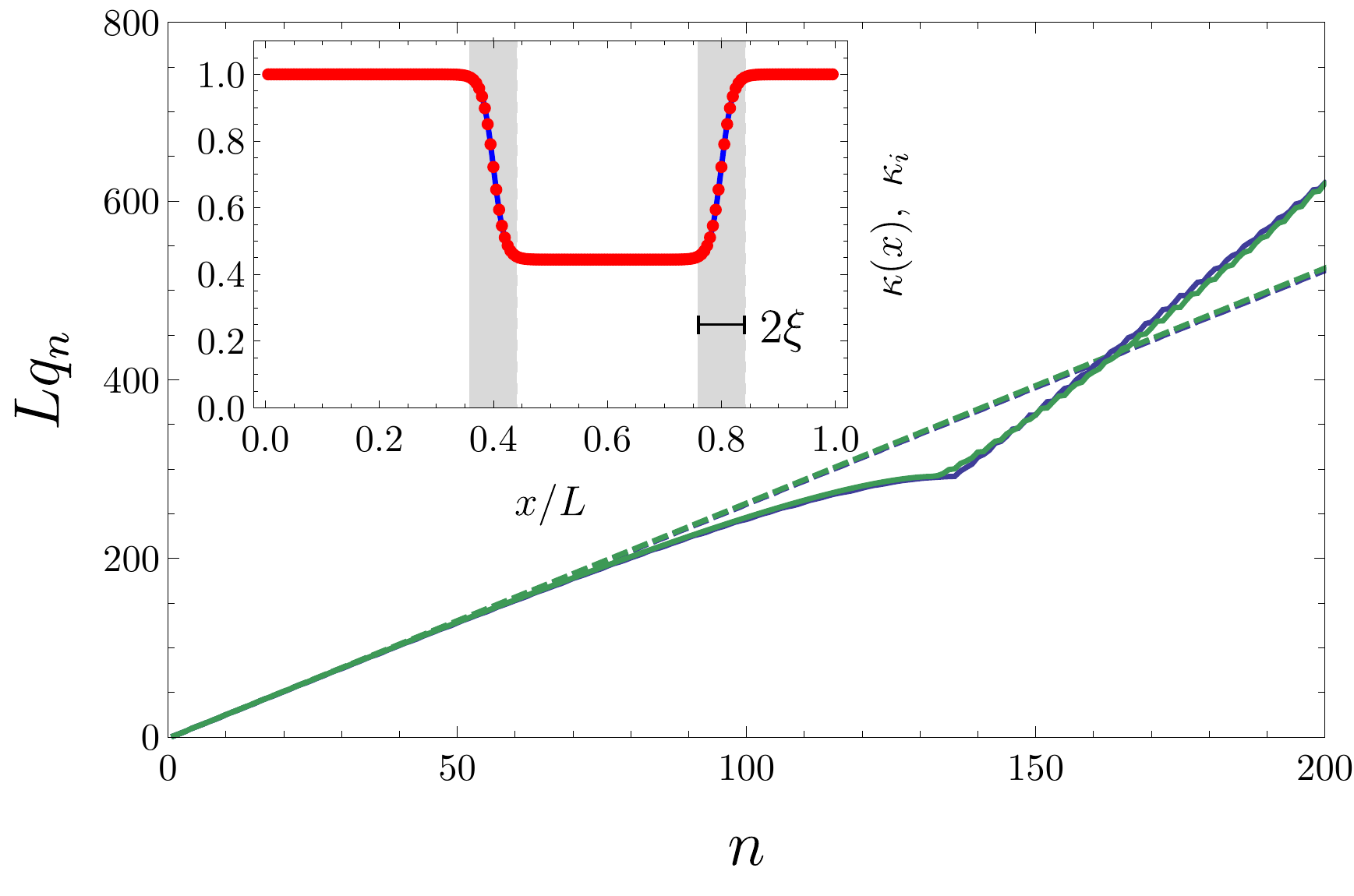}
	\caption{The spectrum of the SFT operator (dashed lines) and the discrete operator (solid lines) for the case $\xi\!=\!L/25$ (green) and $\xi=0$ (blue) which is discussed in the manuscript. Inset: The smoothed $\kappa(x)$ (solid line) and the discrete $\kappa_i$ (points). The parameters used are the same as those of Fig.~\ref{fig:modes} of the main text, together with $\xi=L/25$. The shaded regions, each of width $2\xi$, show the region where $\kappa$ varies.}
	\label{fig:smooth}
\end{figure}

The continuum eigenmodes of a system with a smoothly varying $\kappa(x)$ can be obtained using a straightforward shooting method. The Sturm-Liouville problem associated with $\L\1$, namely
\begin{equation}
\pd{}{x}\left(\kappa(x) \pd{w}{x}\right)=\lambda\,w(x)
\label{eq:ODE}
\end{equation}
\renewcommand{\=}{\!=\!}
is interpreted as a differential equation which is integrated with the initial conditions $w(0)\=0$ and $w'(0)\=1$. The equation is integrated up to $x\=L$ and the value $w'(L;\lambda)$ is obtained as a function of $\lambda$. The eigenvalues are those $\lambda$ for which $w'(L;\lambda)\=0$. These are found using standard root-finding methods.

To explore the effect of the smoothness of $\kappa(x)$ on the results, we chose a specific form of smoothing. Instead of a sharp step function, defined as
\begin{equation}
\Theta(x)=\begin{cases} 0 &x<0 \\1&x>0\end{cases}\ ,
\label{eq:stepfunc}
\end{equation}
we use a hyperbolic tangent function that varies over a finite lengthscale $\xi$
\begin{equation}
\Theta(x;\xi)=\frac{1}{2}\left[1+\tanh\left(\frac{2x}{\xi}\right)\right]\ .
\label{eq:smooth_H}
\end{equation}
$\xi$ can be significantly larger than the monomeric lengthscale. Equation~\eqref{eq:stepfunc} in recovered in the limit $\xi\!\to \! 0$. An example of a smoothed $\kappa(x)$ with $\xi\=L/25$ is shown in Fig.~\ref{fig:smooth}. The computed spectra are also shown and it is seen that the effect of $\xi$ on the spectrum is small and the qualitative discrepancies between the continuum and discrete theories persist. Moreover, the qualitative discrepancies between the continuum and discrete theories are independent of $\xi$, at least as long as $\xi\!\ll\!L$. 

\section{Half-space Gaussian integrals}
\label{sec:gauss}
The partition function, defined in Eq.~\eqref{eq:Z} of the main text, is a multivariate Gaussian integral over a half space. In this section we calculate such an integral in a general manner, to be used in calculations of $Z\DT$ and $Z\SFT$. 

We want to calculate the integral
\begin{equation}
I(\bm A ,\bm v)=\int_{-\infty}^\infty d^N\! \bm x\, e^{-\frac{1}{2} \bm x^T \bm{Ax}} \Theta\left(b-\bm {v}\cdot \bm x\right)\ .
\label{eq:I}
\end{equation}
That is, the integral of a multivariate Gaussian over the half space defined by $\bm v\cdot\bm x<b$. $\bm v$ is an arbitrary real vector and $\bm A$ is a strictly positive-definite symmetric matrix. We begin with the simpler case where $\bm A$ is diagonal. The generalization for the non-diagonal case will be immediate. The integral is then
\begin{equation*}
I(\bm A ,\bm v)=\int_{-\infty}^\infty d^N\!\bm x\, e^{-\frac{1}{2} \sum_i \lambda_i x_i^2 } \Theta\left(b-\bm {v}\cdot \bm x\right)\ ,
\end{equation*}
where the $\lambda_i$'s are the eigenvalues of $\bm A$. We replace the Heaviside function by the integral identity
\begin{equation*}
\Theta(b-x)=\int_{-\infty}^b\!\!\! \delta(z-x)dz=\int_{-\infty}^b\! \frac{dz}{2\pi}\int_{-\infty}^\infty\!\!\! d\omega\, \exp[i\omega(z-x)]\ ,
\end{equation*}
where two auxiliary variables, $\omega$ and $z$, were introduced. This identity holds for arbitrary $x,b\in\mathbb{R}$. With this replacement, after simple rearrangement the integral is written as 
\begin{align}
I&=\int_{-\infty}^b \frac{dz}{2\pi}\int_{-\infty}^\infty e^{i\omega z} \ \times\\
&\Bigg(\prod_{j=1}^N\int_{-\infty}^\infty d x_j\, \exp\left[-\left( \frac{1}{2}\lambda_j x_j^2+i \omega v_j x_j\right)\right]\Bigg)d\omega \ .
\nonumber
\end{align}
This is a product of Gaussian integrals, for each of which we can use the integral identity
\begin{equation}
\int_{-\infty}^{\infty} \exp\left[-\frac{1}{2}a z^2 \pm i \omega z\right]dz=\sqrt{\frac{2\pi}{a}}\exp\left[-\frac{\omega^2}{2a}\right]\ ,
\label{eq:HS}
\end{equation}
which holds for any $\omega\in\mathbb{C}$ and real $a>0$. Thus,
\begin{align}
I&=\int_{-\infty}^b \frac{dz}{2\pi}\int_{-\infty}^\infty d\omega e^{i\omega z} \prod_{i=1}^N\sqrt{\frac{2\pi}{\lambda_i}}\exp\left[-\frac{\omega ^2 v_i^2}{2 \lambda_i}\right]\\
&=\sqrt{\frac{(2\pi)^N}{\det\bm A}}\int_{-\infty}^b \frac{dz}{2\pi}\int_{-\infty}^\infty d\omega 
e^{-\frac{1}{2}\left(\sum\frac{v_i^2}{\lambda_i}\right)\omega ^2 +i\omega b} \ . \nonumber
\end{align}
The latter is again a Gaussian integral of the form of Eq.~\eqref{eq:HS}, and denoting $D\equiv\sum_i\frac{v_i^2}{\lambda_i}=\bm v^T\bm{A}^{-1}\bm v$ we get
\begin{align}
I&=\sqrt{\frac{(2\pi)^N}{\det\bm A}}\int_{-\infty}^b dz\, \sqrt{\frac{2\pi}{D}}\exp\left[-\frac{z^2}{2D}\right]\ .
\end{align}
The last integral is expressed in terms of the standard error function
\begin{equation}
\erf(z)\equiv\frac{2}{\sqrt\pi}\int_0^z e^{-x^2}dx\ ,
\end{equation}
such that
\begin{equation}
\begin{split}
I(\bm A,\bm v)&=\frac{1}{2}\sqrt{\frac{(2\pi)^N}{\det\bm A}}\left[1+\erf\left(\frac{b}{\sqrt{2D}}\right)\right]\ ,\\   
D&\equiv \bm v^T\bm{A}^{-1}\bm v \ .
\end{split}
\label{eq:half_space}
\end{equation}
This completes the derivation. While this is not necessary for the present needs, we note that the formula \eqref{eq:half_space} is valid also when $\bm A$ is not diagonal. This can be seen by a simple change of variables.

\begin{thebibliography}{28}
\providecommand*{\natexlab}[1]{#1}
\providecommand*{\mciteSetBstSublistMode}[1]{}
\providecommand*{\mciteSetBstMaxWidthForm}[2]{}
\providecommand*{\mciteBstWouldAddEndPuncttrue}
  {\def\EndOfBibitem{\unskip.}}
\providecommand*{\mciteBstWouldAddEndPunctfalse}
  {\let\EndOfBibitem\relax}
\providecommand*{\mciteSetBstMidEndSepPunct}[3]{}
\providecommand*{\mciteSetBstSublistLabelBeginEnd}[3]{}
\providecommand*{\EndOfBibitem}{}
\mciteSetBstSublistMode{f}
\mciteSetBstMaxWidthForm{subitem}
{(\emph{\alph{mcitesubitemcount}})}
\mciteSetBstSublistLabelBeginEnd{\mcitemaxwidthsubitemform\space}
{\relax}{\relax}

\bibitem{Ediger2000}
M.~D.~Ediger, \emph{Annu. Rev. Phys. Chem.}, 2000, \textbf{51}, 99--128\relax
\mciteBstWouldAddEndPuncttrue
\mciteSetBstMidEndSepPunct{\mcitedefaultmidpunct}
{\mcitedefaultendpunct}{\mcitedefaultseppunct}\relax
\EndOfBibitem
\bibitem[Cavagna(2009)]{Cavagna2009}
A.~Cavagna, \emph{Phys. Rep.}, 2009, \textbf{476}, 51--124\relax
\mciteBstWouldAddEndPuncttrue
\mciteSetBstMidEndSepPunct{\mcitedefaultmidpunct}
{\mcitedefaultendpunct}{\mcitedefaultseppunct}\relax
\EndOfBibitem
\bibitem{Berthier2011}
L.~Berthier and G.~Biroli, \emph{Rev. Mod. Phys.}, 2011, \textbf{83}, 587--645\relax
\mciteBstWouldAddEndPuncttrue
\mciteSetBstMidEndSepPunct{\mcitedefaultmidpunct}
{\mcitedefaultendpunct}{\mcitedefaultseppunct}\relax
\EndOfBibitem
\bibitem[{De Gennes}(1979)]{deGennes1979}
P.-G. {De Gennes}, \emph{{Scaling concepts in polymer physics}}, Cornell
  university press, 1979\relax
\mciteBstWouldAddEndPuncttrue
\mciteSetBstMidEndSepPunct{\mcitedefaultmidpunct}
{\mcitedefaultendpunct}{\mcitedefaultseppunct}\relax
\EndOfBibitem
\bibitem[Safran(1994)]{Safran1994}
S.~A. Safran, \emph{{Statistical thermodynamics of surfaces, interfaces, and
  membranes}}, Addison-Wesley, 1994\relax
\mciteBstWouldAddEndPuncttrue
\mciteSetBstMidEndSepPunct{\mcitedefaultmidpunct}
{\mcitedefaultendpunct}{\mcitedefaultseppunct}\relax
\EndOfBibitem
\bibitem[Helfand(1975)]{Helfand1975}
E.~Helfand, \emph{J. Chem. Phys.}, 1975, \textbf{62}, 999\relax
\mciteBstWouldAddEndPuncttrue
\mciteSetBstMidEndSepPunct{\mcitedefaultmidpunct}
{\mcitedefaultendpunct}{\mcitedefaultseppunct}\relax
\EndOfBibitem
\bibitem[Popov and Tkachenko(2007)]{Popov2007}
Y.~O. Popov and A.~V. Tkachenko, \emph{Phys. Rev. E}, 2007, \textbf{76},
  021901\relax
\mciteBstWouldAddEndPuncttrue
\mciteSetBstMidEndSepPunct{\mcitedefaultmidpunct}
{\mcitedefaultendpunct}{\mcitedefaultseppunct}\relax
\EndOfBibitem
\bibitem[Su and Purohit(2010)]{Su2010}
T.~Su and P.~K. Purohit, \emph{J. Mech. Phys. Solids}, 2010, \textbf{58},
  164--186\relax
\mciteBstWouldAddEndPuncttrue
\mciteSetBstMidEndSepPunct{\mcitedefaultmidpunct}
{\mcitedefaultendpunct}{\mcitedefaultseppunct}\relax
\EndOfBibitem
\bibitem[Prochniewicz \emph{et~al.}(2005)Prochniewicz, Janson, Thomas, and {De
  la Cruz}]{Prochniewicz2005}
E.~Prochniewicz, N.~Janson, D.~D. Thomas and E.~M. {De la Cruz}, \emph{J. Mol.
  Biol.}, 2005, \textbf{353}, 990--1000\relax
\mciteBstWouldAddEndPuncttrue
\mciteSetBstMidEndSepPunct{\mcitedefaultmidpunct}
{\mcitedefaultendpunct}{\mcitedefaultseppunct}\relax
\EndOfBibitem
\bibitem[McCullough \emph{et~al.}(2008)McCullough, Blanchoin, Martiel, and {De
  La Cruz}]{McCullough2008}
B.~R. McCullough, L.~Blanchoin, J.-L. Martiel and E.~M. {De La Cruz}, \emph{J.
  Mol. Biol.}, 2008, \textbf{381}, 550--8\relax
\mciteBstWouldAddEndPuncttrue
\mciteSetBstMidEndSepPunct{\mcitedefaultmidpunct}
{\mcitedefaultendpunct}{\mcitedefaultseppunct}\relax
\EndOfBibitem
\bibitem[Yogurtcu \emph{et~al.}(2012)Yogurtcu, Kim, and Sun]{Yogurtcu2012}
O.~N. Yogurtcu, J.~S. Kim and S.~X. Sun, \emph{Biophys. J.}, 2012,
  \textbf{103}, 719--27\relax
\mciteBstWouldAddEndPuncttrue
\mciteSetBstMidEndSepPunct{\mcitedefaultmidpunct}
{\mcitedefaultendpunct}{\mcitedefaultseppunct}\relax
\EndOfBibitem
\bibitem[Galkin \emph{et~al.}(2011)Galkin, Orlova, Kudryashov, Solodukhin,
  Reisler, Schr{\"{o}}der, and Egelman]{Galkin2011}
V.~E. Galkin, A.~Orlova, D.~S. Kudryashov, A.~Solodukhin, E.~Reisler, G.~F.
  Schr{\"{o}}der and E.~H. Egelman, \emph{Proc. Natl. Acad. Sci. U. S. A.},
  2011, \textbf{108}, 20568--72\relax
\mciteBstWouldAddEndPuncttrue
\mciteSetBstMidEndSepPunct{\mcitedefaultmidpunct}
{\mcitedefaultendpunct}{\mcitedefaultseppunct}\relax
\EndOfBibitem
\bibitem[Panyukov and Rabin(1996)]{Panyukov1996}
S.~Panyukov and Y.~Rabin, \emph{Macromolecules}, 1996, \textbf{29},
  7960--7975\relax
\mciteBstWouldAddEndPuncttrue
\mciteSetBstMidEndSepPunct{\mcitedefaultmidpunct}
{\mcitedefaultendpunct}{\mcitedefaultseppunct}\relax
\EndOfBibitem
\bibitem[Bensimon \emph{et~al.}(1998)Bensimon, Dohmi, and Mezard]{Bensimon1998}
D.~Bensimon, D.~Dohmi and M.~Mezard, \emph{Europhys. Lett.}, 1998,
  \textbf{42}, 97\relax
\mciteBstWouldAddEndPuncttrue
\mciteSetBstMidEndSepPunct{\mcitedefaultmidpunct}
{\mcitedefaultendpunct}{\mcitedefaultseppunct}\relax
\EndOfBibitem
\bibitem[Lipfert \emph{et~al.}(2010)Lipfert, Klijnhout, and
  Dekker]{Lipfert2010}
J.~Lipfert, S.~Klijnhout and N.~H. Dekker, \emph{Nucleic Acids Res.}, 2010,
  \textbf{38}, 7122--7132\relax
\mciteBstWouldAddEndPuncttrue
\mciteSetBstMidEndSepPunct{\mcitedefaultmidpunct}
{\mcitedefaultendpunct}{\mcitedefaultseppunct}\relax
\EndOfBibitem
\bibitem[Kostjukov and Evstigneev(2012)]{Kostjukov2012}
V.~V. Kostjukov and M.~P. Evstigneev, \emph{Phys. Rev. E.}, 2012, \textbf{86}, 86--88\relax
\mciteBstWouldAddEndPuncttrue
\mciteSetBstMidEndSepPunct{\mcitedefaultmidpunct}
{\mcitedefaultendpunct}{\mcitedefaultseppunct}\relax
\EndOfBibitem
\bibitem[Helfrich(1973)]{Helfrich1973}
W.~Helfrich, \emph{Z. Naturforsch. C}, 1973, \textbf{28},
  693--703\relax
\mciteBstWouldAddEndPuncttrue
\mciteSetBstMidEndSepPunct{\mcitedefaultmidpunct}
{\mcitedefaultendpunct}{\mcitedefaultseppunct}\relax
\EndOfBibitem
\bibitem[Kardar(2007)]{KardarFields}
M.~Kardar, \emph{{Statistical physics of fields}}, Cambridge University Press,
  2007\relax
\mciteBstWouldAddEndPuncttrue
\mciteSetBstMidEndSepPunct{\mcitedefaultmidpunct}
{\mcitedefaultendpunct}{\mcitedefaultseppunct}\relax
\EndOfBibitem
\bibitem[McCullough \emph{et~al.}(2011)McCullough, Grintsevich, Chen, Kang,
  Hutchison, Henn, Cao, Suarez, Martiel, Blanchoin, Reisler, and {De La
  Cruz}]{McCullough2011}
B.~R. McCullough, E.~E. Grintsevich, C.~K. Chen, H.~Kang, A.~L. Hutchison,
  A.~Henn, W.~Cao, C.~Suarez, J.-L. Martiel, L.~Blanchoin, E.~Reisler and E.~M.
  {De La Cruz}, \emph{Biophys. J.}, 2011, \textbf{101}, 151--9\relax
\mciteBstWouldAddEndPuncttrue
\mciteSetBstMidEndSepPunct{\mcitedefaultmidpunct}
{\mcitedefaultendpunct}{\mcitedefaultseppunct}\relax
\EndOfBibitem
\bibitem[Arfken(2013)]{Arfken}
G.~B. Arfken, \emph{{Mathematical methods for physicists}}, Academic press, 4th
  edn, 2013, p. 539\relax
\mciteBstWouldAddEndPuncttrue
\mciteSetBstMidEndSepPunct{\mcitedefaultmidpunct}
{\mcitedefaultendpunct}{\mcitedefaultseppunct}\relax
\EndOfBibitem
\bibitem{SM}
Supplementary material is attached at the end of this file\relax
\mciteBstWouldAddEndPuncttrue
\mciteSetBstMidEndSepPunct{\mcitedefaultmidpunct}
{\mcitedefaultendpunct}{\mcitedefaultseppunct}\relax
\EndOfBibitem
\bibitem[Ashcroft and Mermin(2011)]{AshcroftMermin}
N.~W. Ashcroft and N.~D. Mermin, \emph{{Solid State Physics}}, Cengage
  Learning, 2011\relax
\mciteBstWouldAddEndPuncttrue
\mciteSetBstMidEndSepPunct{\mcitedefaultmidpunct}
{\mcitedefaultendpunct}{\mcitedefaultseppunct}\relax
\EndOfBibitem
\bibitem[Rubinstein and Colby(2003)]{Rubinstein2003}
M.~Rubinstein and R.~H. Colby, \emph{{Polymer Physics}}, Oxford University
  Press, 2003\relax
\mciteBstWouldAddEndPuncttrue
\mciteSetBstMidEndSepPunct{\mcitedefaultmidpunct}
{\mcitedefaultendpunct}{\mcitedefaultseppunct}\relax
\EndOfBibitem
\bibitem[Goulian \emph{et~al.}(1993)Goulian, Bruinsma, and Pincus]{Goulian1993}
M.~Goulian, R.~Bruinsma and P.~Pincus, \emph{Europhys. Lett.}, 1993,
  \textbf{22}, 145--150\relax
\mciteBstWouldAddEndPuncttrue
\mciteSetBstMidEndSepPunct{\mcitedefaultmidpunct}
{\mcitedefaultendpunct}{\mcitedefaultseppunct}\relax
\EndOfBibitem
\bibitem[Golestanian \emph{et~al.}(1996)Golestanian, Goulian, and
  Kardar]{Golestanian1996}
R.~Golestanian, M.~Goulian and M.~Kardar, \emph{Phys. Rev. E}, 1996,
  \textbf{54}, 6725--6734\relax
\mciteBstWouldAddEndPuncttrue
\mciteSetBstMidEndSepPunct{\mcitedefaultmidpunct}
{\mcitedefaultendpunct}{\mcitedefaultseppunct}\relax
\EndOfBibitem
\bibitem[Kardar and Golestanian(1999)]{Kardar1999}
M.~Kardar and R.~Golestanian, \emph{Rev. Mod. Phys.}, 1999, \textbf{71},
  1233--1245\relax
\mciteBstWouldAddEndPuncttrue
\mciteSetBstMidEndSepPunct{\mcitedefaultmidpunct}
{\mcitedefaultendpunct}{\mcitedefaultseppunct}\relax
\EndOfBibitem
\bibitem[Simpson and Leonhardt(2015)]{Simpson2015}
W.~M. Simpson and U.~Leonhardt, \emph{{Forces of the quantum vacuum: An
  Introduction to Casimir Physics}}, World Scientific, 2015\relax
\mciteBstWouldAddEndPuncttrue
\mciteSetBstMidEndSepPunct{\mcitedefaultmidpunct}
{\mcitedefaultendpunct}{\mcitedefaultseppunct}\relax
\EndOfBibitem
\bibitem[Casimir(1948)]{Casimir1948}
H.~B.~G. Casimir, Proc. K. Ned. Akad. Wet., 1948, pp. 793--795\relax
\mciteBstWouldAddEndPuncttrue
\mciteSetBstMidEndSepPunct{\mcitedefaultmidpunct}
{\mcitedefaultendpunct}{\mcitedefaultseppunct}\relax
\EndOfBibitem
\end{thebibliography}
\end{document}